\documentclass[lettersize,journal]{IEEEtran}
\usepackage{amsmath,amsfonts}
\usepackage{array}
\usepackage[caption=false,font=normalsize,labelfont=sf,textfont=sf]{subfig}
\usepackage{textcomp}
\usepackage{stfloats}
\usepackage{url}
\usepackage{verbatim}
\usepackage{graphicx}
\usepackage{tikz}

\usepackage{algorithm}
\usepackage{algpseudocode}
\usepackage{amssymb}
\usepackage{float}

\usetikzlibrary{arrows.meta, positioning, shapes.geometric,calc}
\def\BibTeX{{\rm B\kern-.05em{\sc i\kern-.025em b}\kern-.08em
    T\kern-.1667em\lower.7ex\hbox{E}\kern-.125emX}}
\usepackage{balance}
\usepackage[hidelinks]{hyperref}
\begin{document}
\title{LEAD: A Local Ensemble-Assisted Parallel Decoding Framework for Quantum Tanner Codes}
\author{~Zhuo-Yan Xiao,
       ~Sha~Shi,~Chen-Peng Huang,~Dong-Sheng Wang,
        and Yun-Jiang~Wang~\IEEEmembership{Member,~IEEE,}
\thanks{
ZYX, SS, CPH and YJW are all with the State key laboratory of Integrated Services Networks and school of telecommunication engineering, Xidian University, Xi'an, Shann Xi 710071, China. DSW is with the Institute of Theoretical Physics, Chinese Academy of Sciences, Beijing 100190, China.
e-mail: (zyxiao-1@stu.xidian.edu.cn; sshi@xidian.edu.cn; fowardns@163.com; wds@itp.ac.cn; yunjiangw@xidian.edu.cn  )}}

\markboth{}
{How to Use the IEEEtran \LaTeX \ Templates}

\maketitle

\begin{abstract}
Quantum Tanner codes are a recently developed family of quantum error-correcting codes characterized by favorable asymptotic performance characteristics. Despite their theoretical potential, practical decoding algorithms that effectively leverage their structural properties remain limited. This work introduces LEAD (Local Ensemble-Assisted Decoder), a structure-aware decoding framework tailored for quantum Tanner codes. The proposed scheme leverages the decomposable structure of Cayley complexes to project the global code onto overlapping local subcodes defined by vertex neighborhoods, where error probabilities are estimated in parallel. To ensure global consistency, LEAD utilizes the inherent topological symmetry of the complex and introduces a soft-information regularization mechanism to mitigate local overconfidence during information aggregation. This framework enables highly parallelized, low-complexity decoding that is intrinsically compatible with various local search heuristics. Simulation results demonstrate that LEAD achieves significantly lower logical error rates than standard decoding framework while substantially reducing the average decoding latency and iteration count.

\end{abstract}

\begin{IEEEkeywords}
Quantum error correction, quantum low-density parity-check (qLDPC) codes, quantum Tanner codes, belief propagation
\end{IEEEkeywords}

\section{Introduction}
Quantum computing offers capabilities that surpass those of classical computers in tasks such as integer factorization and quantum simulation \cite{cornish2024quantum}. Nevertheless, quantum states are highly vulnerable to noise, which necessitates the use of quantum error-correcting codes (QECCs) for fault-tolerant computation \cite{campbell2017roads,gottesman2013fault}.
Among various approaches, quantum low-density parity-check (QLDPC) codes have attracted considerable attention due to their sparse stabilizer structure and relatively low overhead in auxiliary qubits \cite{babar2015fifteen}.

Recent advances in QLDPC code design have led to constructions whose parameters approach those of asymptotically good codes \cite{breuckmann2021balanced,hastings2021fiber,panteleev2021quantum}.
Building on this progress, Leverrier and Zémor introduced quantum Tanner codes, a framework for constructing asymptotically good quantum LDPC codes.
These codes achieve a minimum distance that scales linearly with code length, enabling the correction of a proportionally growing number of errors \cite{leverrier2022quantum,gu2023efficient,leverrier2023decoding,mostad2024generalizing}.
Recently constructed explicit quantum Tanner codes \cite{leverrier2025small, radebold2025explicit} exhibit favorable finite-size performance and competitive pseudo-thresholds under both phenomenological and circuit-level noise \cite{radebold2025explicit}, yielding lower space-time overhead than traditional surface codes.
These developments suggest that quantum Tanner codes may offer practical advantages even at short code lengths, thereby underscoring the need for efficient decoding algorithms.

Theoretical breakthroughs by Gu et al. \cite{gu2023efficient} and Leverrier $\&$ Zémor \cite{leverrier2023decoding} have established that quantum Tanner codes can be decoded in linear time up to a constant fraction of the code length. These algorithms rely on minimizing locally defined cost functions or mismatch vectors within specific neighborhoods. However, their practical implementation is hindered by a worst-case computational complexity that grows exponentially with the size of the local code or constraint support. This exponential scaling, combined with the need for exhaustive local enumerations, significantly restricts their scalability and performance for practical block lengths and moderate constraint weights.

Belief propagation (BP) remains the foundational decoding algorithm for LDPC codes \cite{kschischang2002factor}. In the quantum domain, however, its efficacy is severely hindered by the prevalence of short cycles and the high degeneracy of quantum codes. To address these challenges, current research largely relies on post-processing strategies, most notably Ordered Statistics Decoding (OSD) and the more recent Localized Statistical Decoding (LSD) \cite{hillmann2025localized,panteleev2021degenerate}.
While these iterative approaches do not suffer from the prohibitive computational complexity associated with the deterministic decoders proposed by Gu et al. and Leverrier $\&$ Zémor, they are inherently structure-blind when applied to quantum Tanner codes. Specifically, they treat the parity-check matrix as a generic sparse graph and fail to exploit the rich geometric and topological constraints embedded in the underlying Cayley complex.

To bridge this gap between general iterative decoding and code-specific structural properties, we propose LEAD (Local Ensemble-Assisted Decoder), a decoding framework tailored for quantum Tanner codes. The framework operates through a synergistic cascade of local and global decoding stages. It first leverages the structural decomposability of Cayley complexes to project the global code onto overlapping local tensor subcodes defined by the local neighborhoods of complex vertices. Each local subcode is then processed in parallel by an efficient local decoder to estimate initial error probabilities. Subsequently, by exploiting the inherent topological symmetry of the Cayley complex and introducing a soft-information regularization mechanism to mitigate local overconfidence, these local estimates are effectively aggregated into a refined global error probability prior. This prior is finally utilized by a global decoder to achieve a consistent and accurate error correction.

To ensure a fair comparison, we employ consistent BP and OSD/LSD parameters across both the LEAD framework and the standard baselines. The key distinction lies in the decoding topology: while conventional baselines operate directly on the global parity-check matrix, LEAD deconstructs the global decoding problem into parallelized, structure-informed local clusters. Simulation results demonstrate that LEAD significantly outperforms these conventional, structure-blind decoders. Although LEAD utilizes standard algorithms as its local sub-decoders, its hierarchical architecture enables superior error localization and suppression compared to a direct global approach. Crucially, the generation of highly accurate soft priors within the local stage substantially reduces the iteration count required by the global decoder, thereby enhancing total decoding throughput.

Furthermore, at short and moderate code lengths, LEAD offers greater practicality than existing theoretical local-decoder-based methods \cite{gu2023efficient, leverrier2023decoding} by obviating the need for computationally expensive mismatch-vector computations. By offloading intensive computational tasks to parallelized local clusters, LEAD maintains a low-latency profile, effectively translating the code's theoretical potential into tangible practical performance, providing a robust and extensible platform for future integration with advanced soft-output heuristics.

The remainder of this paper is organized as follows.
Section~\ref{sec:background} reviews the fundamentals of stabilizer codes
and outlines the construction of quantum Tanner codes.
Section~\ref{sec:lead} introduces the proposed LEAD framework and provides a detailed description of its decoding procedure.
Section~\ref{sec:results} presents simulation results and performance comparisons against BP-OSD and BP-LSD decoder.
Section~\ref{sec:discussion} discusses the relationship with concurrent works and explores future directions.
Finally, Section~\ref{sec:conclusion} concludes the paper and summarizes the main contributions.

\section{Background and Code Constructions}\label{sec:background}

In classical coding theory, the code space of a linear code corresponds to the null space of its parity-check matrix. In quantum error correction, stabilizer codes play an analogous role to classical linear codes.
The valid code space of an $n$-qubit stabilizer code is defined as the simultaneous $+1$ eigenspace of all $m$ independent generators of an Abelian subgroup $S$, where
\begin{equation}
S = \langle S_1, \dots, S_m \rangle \subseteq \mathcal{G}_n.
\end{equation}
Here, $\mathcal{G}_n$ denotes the $n$-qubit Pauli group, whose elements are formed by the $n$-fold tensor product of single-qubit Pauli operators $\sigma \in \{I, X, Y, Z\}$, where $Y = iXZ$.
All generators in $S$ are required to mutually commute and $-I \notin S$ to ensure a non-trivial code space.
Furthermore, if the stabilizer group can be generated by a set of operators involving only Pauli $X$ or only Pauli $Z$, the corresponding code is referred to as a Calderbank–Shor–Steane (CSS) code \cite{steane1996error,calderbank1996good}.
Most quantum error-correcting codes studied in the literature belong to this CSS class.

Quantum Tanner codes are a class of CSS codes and can be viewed as a high-dimensional generalization of classical Tanner codes.
They are constructed from a high-dimensional graph structure known as the left–right Cayley complex \cite{leverrier2022quantum}.
In this paper, we first analyze the left–right Cayley complex used in the construction of quantum Tanner codes, and then briefly review the overall construction procedure.
For a more detailed account of quantum Tanner codes, readers are referred to the original work of Leverrier and Zémor \cite{leverrier2022quantum}.

\subsection{The Left-Right Cayley Complex}

The left–right Cayley complex can be constructed from a finite group $G$ together with two subsets of generators $A, B \subseteq G$, where $A = A^{-1}$ and $B = B^{-1}$. The cardinalities of $A$ and $B$ need not be equal; however, for simplicity and without loss of generality, this paper considers only the case $|A| = |B| = \Delta$ \cite{leverrier2022quantum,gu2023efficient,leverrier2023decoding}.

\begin{figure}
    \centering
    \includegraphics[width=0.6\linewidth]{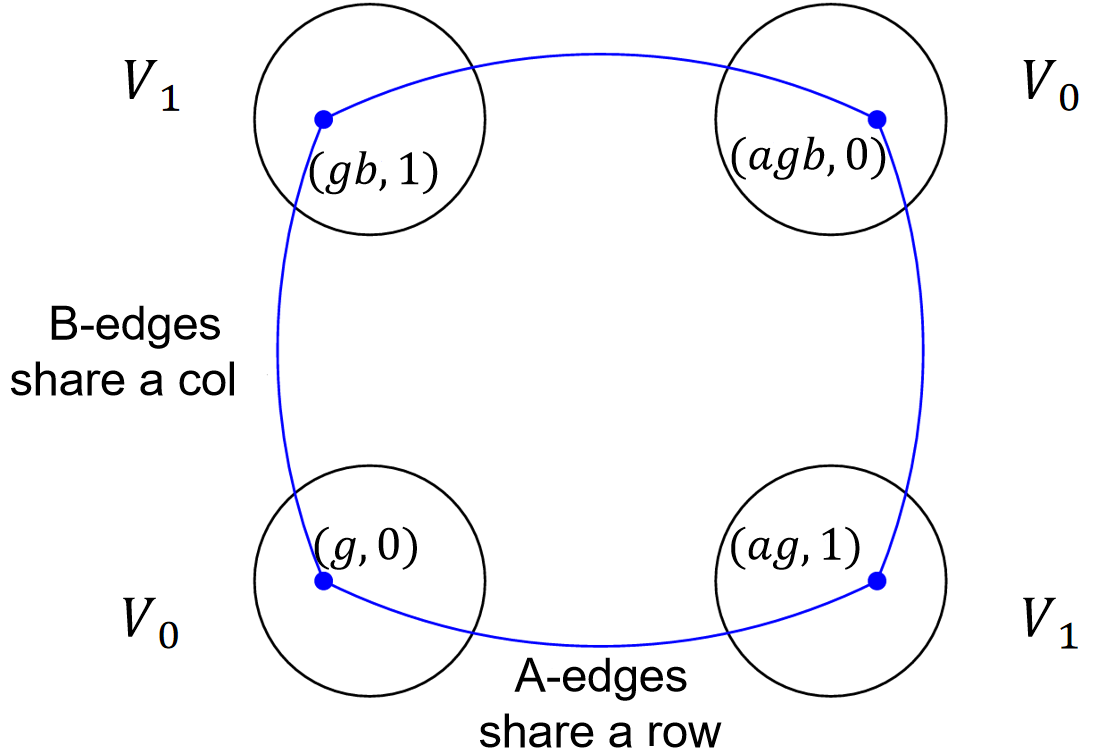}
    \caption{As described in the main text, each face is composed of four edges, where the horizontal edges ($A$-edges) and vertical edges ($B$-edges) correspond to the group elements $a \in A$ and $b \in B$, respectively \cite{leverrier2022quantum}.
}
    \label{fig:cayley}
\end{figure}

The left-right Cayley complex $X$ is an incidence structure between a set of vertices, two sets of edges, and a set of squares. The construction is detailed as follows:

\begin{enumerate}
\item \textbf{Vertex Set}:
The vertex set $V$ is partitioned as $V = V_0 \cup V_1$, where both $V_0$ and $V_1$ are identified as copies of the group $G$. Formally, $V_0 = G \times \{0\}$ and $V_1 = G \times \{1\}$.

\item \textbf{Edge Set}:
The edge set $E$ is partitioned into $A$-edges ($E_A$) and $B$-edges ($E_B$). A vertex $(g,0) \in V_0$ and a vertex $(g',1) \in V_1$ are connected by an $A$-edge if $g' = ag$ for some $a \in A$, and by a $B$-edge if $g' = gb$ for some $b \in B$. The graphs $\mathcal{G}_A = (V, E_A)$ and $\mathcal{G}_B = (V, E_B)$ are the double covers of the left and right Cayley graphs $\mathrm{Cay}(G,A)$ and $\mathrm{Cay}(G,B)$, respectively.

\item \textbf{Face Set}:
The set of faces $Q$ consists of quadrangles (squares) on which the physical qubits are placed. As illustrated in Fig.~\ref{fig:cayley}, each face is defined as a 4-subset of vertices:
\end{enumerate}
\begin{equation}
\label{eq:group_set}
\{ (g,0),\ (ag,1),\ (gb,1),\ (agb,0) \}, \quad g \in G, a \in A, b \in B
\end{equation}
 To ensure each square contains exactly four distinct vertices and that every vertex is incident to exactly $\Delta^2$ squares, the sets $A$ and $B$ must satisfy the Total No-Conjugacy (TNC) condition: $ag \neq gb$ for all $a \in A, b \in B, g \in G$. This condition implies that the $Q$-neighborhood of any vertex is in bijection with the product set $A \times B$.

\subsection{Quantum Tanner Code  Construction and Local Constraints}
The Quantum Tanner code is defined by assigning physical qubits to the set of faces $Q$ and enforcing local constraints on the links of each vertex. Utilizing the aforementioned bijection $\phi_v: Q(v) \to A \times B$ , the local view of the qubits at each vertex is mapped to a $\Delta \times \Delta$ array. We then associate two classical linear codes, $C_A$ and $C_B$ of length $\Delta$, to define the stabilizer generators as follows:

Z-type Generators: Associated with $V_0$, where the local view $x_v$ must belong to the code $C_0 = C_A \otimes C_B$.

X-type Generators: Associated with $V_1$, where the local view $x_v$ must belong to the code $C_1 = C_A^{\perp} \otimes C_B^{\perp}$.

Equivalently, the quantum code $\mathcal{Q} = (\mathcal{C}_0, \mathcal{C}_1)$ is defined as a pair of Tanner codes $\mathcal{C}_0 = T(\mathcal{G}_0^{\square}, C_0^{\perp})$ and $\mathcal{C}_1 = T(\mathcal{G}_1^{\square}, C_1^{\perp})$. This construction ensures the CSS condition $H_X H_Z^T = 0$ as the generators are orthogonal by design. The resulting code is inherently LDPC, with generators of weight at most $\Delta^2$.

\section{Proposed LEAD Decoder}\label{sec:lead}

In this section, we present the LEAD framework designed for quantum Tanner codes. Given that quantum Tanner codes are CSS-type codes, we focus our discussion on $Z$-type errors, with the decoding of $X$-type errors following by symmetry.

\subsection{Practical Challenges of Theoretical Local-Decoding}

Recent decoding algorithms for quantum Tanner codes attempt to exploit their local structure.
In particular, \cite{leverrier2023decoding} proposed a decoder based on mismatch vectors, which measure inconsistencies among local decoding views defined on the Tanner complex.
The algorithm iteratively reduces the Hamming weight of the mismatch vector through a mismatch decomposition procedure, thereby progressively refining the decoding result.

While this approach provides an elegant theoretical framework, it faces several practical challenges.
First, the decoding process implicitly assumes that the local codes can be reliably resolved, which may not hold even for short or moderate block lengths under realistic noise conditions.
Second, the mismatch decomposition step requires searching over local codewords belonging to product-code structures such as
$C_A^\perp \otimes \mathbb{F}_2^{B}$ and $\mathbb{F}_2^{A} \otimes C_B^\perp$.
By exhaustive search, the corresponding spaces have sizes
$2^{k_A^\perp |B|}$ and $2^{|A| k_B^\perp}$, respectively, which grow exponentially with the sizes of the generator sets $|A|$ and $|B|$.
As a result, the worst-case computational complexity of mismatch decomposition increases rapidly as the local constraint weight grows.

These limitations motivate the development of alternative decoding strategies that can leverage the local structural information of quantum Tanner codes while maintaining low computational complexity.

\begin{figure}[htbp]
\centering
\begin{tikzpicture}[
    node distance=0.8cm,
    module/.style={draw, rectangle, minimum width=1.5cm, minimum height=0.8cm, font=\small},
    arrow/.style={-Stealth, semithick},
    line/.style={semithick},
    merge/.style={circle, draw, inner sep=1pt, minimum size=18pt, font=\footnotesize}
]

\node[module] (dec1) at (2,0) {$\text{DEC}_{v_1}$};
\node[module] (dec2) at (2,-1.5) {$\text{DEC}_{v_2}$};
\node (dots) at (2,-2.25) {$\vdots$};
\node[module] (deck) at (2,-3) {$\text{DEC}_{v_{|V_0|}}$};
\node[module] (ml) at (6.5,-1.5) {G\_DEC};

\node[merge] (merge) at (4.5,-1.5) {$\alpha\Sigma$};

\draw[line] (-0.5,-1.5) -- (0,-1.5);
\draw[line] (0,-3) -- (0,0);
\draw[arrow] (0,0) -- (dec1.west);
\draw[arrow] (0,-1.5) -- (dec2.west);
\draw[arrow] (0,-3) -- (deck.west);

\node[anchor=south] at (-0.25,-1.5) {$\mathbf{s}$};
\node[anchor=south] at (0.9,0) {$\mathbf{s}_{v_1}$};
\node[anchor=south] at (0.9,-1.5) {$\mathbf{s}_{v_2}$};
\node[anchor=south] at (0.9,-3) {$\mathbf{s}_{v_{|V_0|}}$};

\node[anchor=south] at (3.25,0) {$\hat{\mathbf{p}}_{v_1}$};
\node[anchor=south] at (3.25,-1.5) {$\hat{\mathbf{p}}_{v_2}$};
\node[anchor=south] at (3.25,-3) {$\hat{\mathbf{p}}_{v_{|V_0|}}$};

\node[anchor=south] at (5.5,-1.5) {$\hat{\mathbf{p}}$};

\draw[arrow] (dec1.east) -| (merge.north);
\draw[arrow] (dec2.east) -- (merge.west);
\draw[arrow] (deck.east) -| (merge.south);
\draw[arrow] (merge.east) -- (ml.west);

\draw[arrow] (ml.east) -- (8,-1.5);
\node[anchor=west] at (8,-1.5) {$\hat{\mathbf{e}}$};

\end{tikzpicture}
\caption{Flowchart of the LEAD decoding algorithm.}
\label{fig:decoder_parallel}
\end{figure}
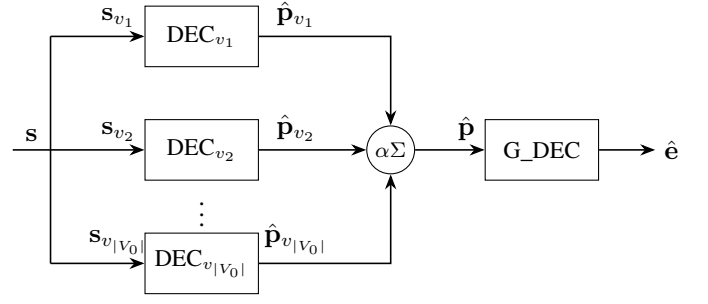
\subsection{Overview of the LEAD Framework}

To effectively exploit the local structural information of quantum Tanner codes while maintaining low complexity, we propose a local-decoder-based framework for quantum decoding, namely the LEAD algorithm.
Unlike mismatch-based approaches that explicitly search for local codewords, LEAD exploits the local tensor-code structure of quantum Tanner codes to estimate channel error probabilities, which are subsequently utilized by a global decoder. The key idea is to decompose the decoding task into two stages.
First, the local tensor codes defined on the $Q$-neighborhoods of vertices in the Cayley complex are decoded independently using efficient local decoders.
These local decoders produce estimates of the error probabilities associated with qubits in their respective neighborhoods.
Second, the locally estimated probabilities are aggregated to form a global prior, which is then provided to a global decoder to obtain the final error estimate.

The workflow of the proposed algorithm is illustrated in Fig.~\ref{fig:decoder_parallel}, where the input is the global syndrome $\mathbf{s}$ and the output is the estimated error $\hat{\mathbf{e}} \in \mathbb{F}_2^{n}$.
Here, the modules $\text{DEC}_{v_1}, \dots, \text{DEC}_{v_{|V_0|}}$ denote the local decoders operating on the respective local views for each vertex $v \in V_0$, while $\text{G\_DEC}$ denotes the global decoder.
The vectors $\mathbf{s}_{v_1}, \dots, \mathbf{s}_{v_{|V_0|}}$ represent the extracted local syndromes, $\hat{\mathbf{p}}_{v_1}, \dots, \hat{\mathbf{p}}_{v_{|V_0|}}$ denote the locally estimated error probability vectors, and $\hat{\mathbf{p}}$ represents the regularized global error probability vector.
All outputs from the local decoders are aggregated through a merge node, scaled by a scaling parameter $\alpha$ to mitigate local overconfidence, and subsequently fed into the global decoder as prior soft information.
This design fully exploits the parallelism of local codes at different nodes, thereby significantly improving the overall decoding efficiency.
The algorithmic steps are summarized in Algorithm~\ref{alg:lead}, where the decoding procedure strictly consists of three phases: local decoding, soft information regularization, and global decoding, as detailed below.

\begin{algorithm}[!t]
\caption{LEAD Decoder}
\label{alg:lead}
\renewcommand{\algorithmicrequire}{\textbf{Input:}}
\renewcommand{\algorithmicensure}{\textbf{Output:}}

\begin{algorithmic}[1]
\Require Global parity-check matrix $\mathbf{H}$, syndrome $\mathbf{s}$, scaling parameter $\alpha \in (0, 1]$
\Ensure Either failure or an estimated error $\hat{\mathbf{e}} \in \mathbb{F}_2^{n}$ satisfying $\mathbf{H}\hat{\mathbf{e}}=\mathbf{s}$

\State Initialize global prior probability vector $\hat{\mathbf{p}} \leftarrow \mathbf{0}$

\State \textbf{// Phase 1: Local Decoding \& Confidence Boosting}
\For{each vertex $v \in V_0$ }
    \State Extract local parity-check matrix $\mathbf{H}_v$ and local syndrome $\mathbf{s}_v$
    \State $(\hat{\mathbf{p}}_v, \hat{\mathbf{e}}_v) \leftarrow \textsc{LocalDecoder}(\mathbf{H}_v,\mathbf{s}_v)$

    \If{$\mathbf{H}_v \hat{\mathbf{e}}_v = \mathbf{s}_v$} \Comment{If local hard decision succeeds}
        \For{each local bit $j$ where $\hat{e}_{v,j} = 1$}
            \State $\hat{p}_{v,j} \leftarrow \max(\hat{p}_{v,j}, 0.5)$ \Comment{Boost confidence for flipped bits}
        \EndFor
    \EndIf
    \State Record $\hat{\mathbf{p}}_v$ for subsequent aggregation
\EndFor

\State \textbf{// Phase 2: Soft Information Regularization}
\For{each qubit $i \in \{1,\dots,n\}$}
    \State $\mathcal{S}_i \leftarrow \{ \hat{p}_{v,i} : \text{all local views } v \text{ where } i \in Q(v) \}$
    \State $\hat{p}_i \leftarrow \alpha \cdot \textsc{Average}(\mathcal{S}_i)$ \Comment{Aggregate across ALL views and dampen}
\EndFor

\State \textbf{// Phase 3: Global Decoding}
\State $\hat{\mathbf{e}} \leftarrow \textsc{GlobalDecoder}(\mathbf{H},\mathbf{s},\hat{\mathbf{p}})$

\State \Return $\hat{\mathbf{e}}$
\end{algorithmic}
\end{algorithm}

\begin{enumerate}

\item \textbf{Phase 1: Local decoding and confidence boosting.}
For each vertex $v \in V_0$, the decoder extracts the local parity-check matrix $\mathbf{H}_v$ and the corresponding local syndrome $\mathbf{s}_v$.
The local decoder $\textsc{LocalDecoder}(\mathbf{H}_v,\mathbf{s}_v)$ is then invoked to produce a local hard-decision error estimate $\hat{\mathbf{e}}_v$ together with the soft error probability vector $\hat{\mathbf{p}}_v$.
To better exploit valid local structures, a heuristic confidence boost is applied: if the local estimate satisfies the parity-check condition ($\mathbf{H}_v\hat{\mathbf{e}}_v=\mathbf{s}_v$), the soft probabilities of bits predicted to be flipped (i.e., $\hat{e}_{v,j}=1$) are bounded from below by $0.5$.
The resulting probability vector $\hat{\mathbf{p}}_v$ from each local view is then recorded for aggregation.

\item \textbf{Phase 2: Soft-information regularization.}
Rather than discarding failed local estimates, LEAD aggregates soft information across all local views to construct a global prior.
For each qubit $i\in\{1,\dots,n\}$, the locally estimated probabilities $\hat{p}_{v,i}$ from all local views containing qubit $i$ are collected into a set $\mathcal{S}_i$.
To mitigate potential overconfidence (e.g., from false-positive local convergences), the aggregated probability is scaled by a scaling parameter $\alpha\in(0,1]$:
\[
\hat{p}_i=\alpha\cdot\textsc{Average}(\mathcal{S}_i).
\]

\item \textbf{Phase 3: Global decoding.}
The regularized probability vector $\hat{\mathbf{p}}$ is used as updated prior soft information for the global channel.
The global decoder $\textsc{GlobalDecoder}(\mathbf{H},\mathbf{s},\hat{\mathbf{p}})$ is then executed to obtain the final error estimate $\hat{\mathbf{e}}\in\mathbb{F}_2^n$.
If the estimate satisfies the global syndrome equation $\mathbf{H}\hat{\mathbf{e}}=\mathbf{s}$, the decoder outputs $\hat{\mathbf{e}}$; otherwise, a decoding failure is declared.

\end{enumerate}



The scaling parameter $\alpha$ is primarily introduced to regulate the confidence level of the locally estimated priors. While the unscaled configuration ($\alpha = 1.0$) is typically well-calibrated and highly reliable for general quantum Tanner codes, our investigation identifies specific scenarios where such hard local assistance can be counterproductive. As analyzed in the simulation results of Section IV-C, codes with certain expansion properties—such as the $[\![144, 8, 12]\!]$ instance—can trigger a positive feedback loop of erroneous soft information during the global message-passing stage. In these cases, employing $\alpha < 1$ acts as a necessary regularization to suppress local overconfidence and restore global decoding performance (see Fig.~\ref{fig:compare_alpha}).

\subsection{Complexity Analysis of LEAD with BP--OSD}

In this section, we evaluate the computational complexity of the proposed LEAD algorithm, assuming that the local decoders ($\text{DEC}_v$) utilize the BP--LSD algorithm, while the global decoder ($\text{G\_DEC}$) employs the BP--OSD algorithm. According to the construction of quantum Tanner codes over the left-right Cayley complex, the length of the global quantum code is given by $n = |G|\Delta^2/2$, where $|G|$ denotes the order of the finite group $G$ and $\Delta = |A| = |B|$ represents the degree of the underlying Cayley graphs. Noting that the vertex set $V_0$ in the left-right Cayley complex is in one-to-one correspondence with the group elements, we have $|V_0| = |G|$, which determines the total number of parallel local decoders. Correspondingly, each local decoder operates on a specific subcode defined on the $Q$-neighborhood of a vertex, with a subcode length of $n_{\text{loc}} = \Delta^2$.

For BP--LSD decoding, the computational complexity depends on the problem structure (in particular, the size of error clusters).
In the worst case, the complexity can scale as $O(n^3)$; in contrast, under typical conditions with low noise, bounded degree, and bounded error clusters, the overall workload is approximately linear, i.e., $O(n)$. The complexity of the global BP--OSD decoder is dominated by the OSD post-processing stage, which involves Gaussian elimination and information-set reprocessing,the complexity typically scales as $O(n^3)$.
Based on this, the complexity of LEAD can be bounded as follows.

\paragraph{Worst-case upper bound (conservative bound)}
Substituting the code lengths of the local and global decoders into the $O(n^3)$ complexity model, we obtain the total workload as follows:
\begin{align}\label{eq:lead_bplsd_upper}
\underbrace{|G| \cdot O\big((\Delta^2)^3\big)}_{\text{Local Phase}} + \underbrace{O\big((|G|\Delta^2/2)^3\big)}_{\text{Global Phase}}
&= O\big(|G|\Delta^6\big) + O\big(|G|^3\Delta^6\big) \notag\\
&= O\big(n \Delta^4\big) + O\big(n^3\big).
\end{align}
Here, the first term $O(|G|\Delta^6)$ represents the aggregate complexity of the $|G|$ parallel local decoders, each operating on a $Q$-neighborhood of size $\Delta^2$. The second term $O(|G|^3\Delta^6)$ represents the complexity of the global decoder. Given that $n = |G|\Delta^2/2$, the global term dominates the total workload, yielding an overall complexity of $O(n^3)$.

\paragraph{Typical-case lower bound (linear regime)}
Under common settings with low noise, bounded degree, and bounded error clusters, the complexity of BP--LSD scales linearly with the code length. Substituting the respective code lengths into $O(n)$ yields the total computational workload:

\begin{align}
\label{eq:lead_bposd_lower}
\underbrace{|G| \cdot O\big(\Delta^2\big)}_{\text{Local Phase}}
+
\underbrace{O\big((|G|\Delta^2/2)^3\big)}_{\text{Global Phase}}
&=
O\big(|G|\Delta^2\big)
+
O\big(|G|^3\Delta^6\big)
\notag\\
&=
O(n)
+
O(n^3).
\end{align}

This result indicates that, in the high-fidelity regime, the additional computational overhead introduced by the LEAD local decoding stage remains linear with respect to the code length. Consequently, the overall asymptotic complexity continues to be dominated by the global BP--OSD decoding stage.

While the expressions in \eqref{eq:lead_bplsd_upper} and \eqref{eq:lead_bposd_lower} characterize the total computational workload (i.e., the sum of operations across all units), the actual decoding latency can be significantly reduced through hardware parallelism. In a fully parallelized implementation with $|G|$ local processors, the time complexity of the local decoding stage (Phase 1) is determined by the complexity of a single local decoder operating on a constant-sized neighborhood $Q(v)$, i.e., $T_{\text{local}} = O(\Delta^2)$. Since $\Delta$ is a fixed parameter determined by the code construction, $T_{\text{local}} \approx O(1)$ with respect to the global code length $n$. Consequently, the overall parallel runtime of LEAD is dominated by the global decoding stage (Phase 3). By providing high-quality initial priors, LEAD effectively reduces the number of iterations required by the global decoder. Therefore, although the asymptotic complexity order remains similar to that of a standalone global decoder, the practical wall-clock time is substantially diminished. This feature makes LEAD particularly suitable for real-time quantum error correction on specialized hardware such as FPGAs or ASICs.

\paragraph{Summary}

In conclusion, the integration of local decoders in the LEAD framework maintains the overall asymptotic complexity of the underlying algorithm, with the overall complexity remaining dominated by the global BP--OSD decoding stage. Crucially, the local assistance stage serves as a constant-overhead pre-conditioner that significantly reduces the required global iterations. Furthermore, the LEAD architecture is algorithm-agnostic; it can be readily combined with various decoding primitives beyond BP--LSD, providing a versatile and scalable platform for the next generation of quantum Tanner code decoders.

\section{Simulation Results}\label{sec:results}
\subsection{Simulation Setup}

In this section, we evaluate the performance of several quantum Tanner codes with varying code lengths and rates. We compare our proposed LEAD framework—a hierarchical decoding architecture—against two baseline configurations: the standalone application of BP--OSD and BP--LSD directly to the global Tanner complex. While the baselines represent the conventional non-cooperative decoding approach, the LEAD algorithm utilizes a redistributed decoding logic across local and global stages. We quantify the advantages of this proposed architecture in terms of both logical error rates (LER) and normalized decoding iterations.
Both BP--OSD and BP--LSD are implemented using the LDPC Python library \cite{Roffe_LDPC_Python_tools_2022,roffe_decoding_2020}, with the maximum number of iterations set equal to the code length, and the BP component configured to use the min-sum algorithm.
Specifically, the OSD decoder is run in the \texttt{osd\_cs} mode with order 3, while the LSD decoder is run in the \texttt{lsd\_cs} mode, also with order 3.

Within the LEAD framework, we specifically employ BP-LSD as the local sub-decoder and BP-OSD for the global decoding stage. This configuration is strategically chosen to provide reliable local soft priors while maintaining a low-latency profile. As further analyzed in Section \ref{sec:Generalizability Analysis}, this combination effectively balances local error estimation with global convergence stability, particularly in mitigating the local overconfidence phenomenon often associated with more aggressive local decoders. To systematically represent our chosen architecture, we introduce a modular notation: LEAD([Local]--[Global]). Here, BL and BO denote BP--LSD and BP--OSD, making our default configuration LEAD(BL--BO). We quantify the advantages of this proposed architecture in terms of both logical error rates (LER) and normalized decoding iterations.

In the simulations, we adopt the depolarizing noise model with error probability $p$, where each qubit undergoes an $X$, $Y$, or $Z$ error with probability $p/3$ each, and remains unchanged with probability $1-p$.
During decoding, let $\hat{E}$ denote the estimated error and $E$ the actual error. Decoding is successful if $\hat{E}E \in S$, and fails if $\hat{E}E \in N(S)\setminus S$ where $N(S)$ is the normalizer of $S$, in which case the residual operator corresponds to a nontrivial logical operator.

The overall iteration count $I$ of the LEAD algorithm is defined as
\begin{equation}
	\label{iteration}
    I = I_g + I_l \cdot \frac{m_l}{m_g},
\end{equation}
where $I_g$ denotes the number of iterations performed by the global decoder,
$I_l$ denotes the number of iterations performed by the local decoder,
and $m_l$ and $m_g$ are the numbers of parity-check equations (rows of the parity-check matrix) in the local and global codes, respectively. The factor $\frac{m_l}{m_g}$ is introduced to normalize the iteration cost with respect to the number of parity-check constraints involved in each decoding stage. This normalization ensures that the computational cost of a single local iteration is appropriately weighted relative to a global iteration, reflecting the smaller constraint support size of the local subcodes relative to the overall Tanner complex.

\subsection{Random Quantum Tanner Codes}

\begin{figure*}[t!]
    \centering
    \subfloat[]{
        \includegraphics[width=0.3\linewidth]{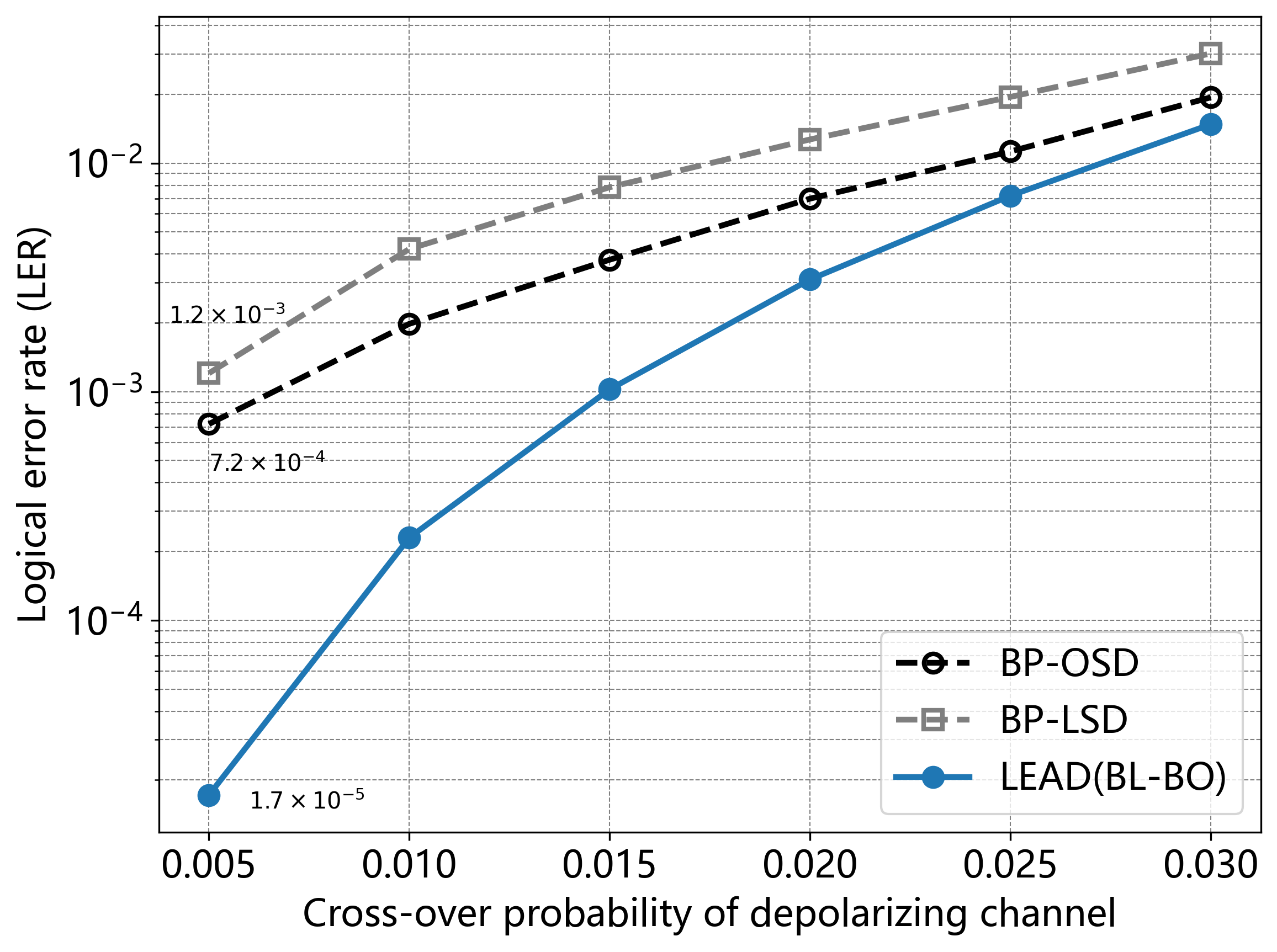}
        \label{fig:Hamming(7,4)}
    }
    \hfill
    \subfloat[]{
        \includegraphics[width=0.3\linewidth]{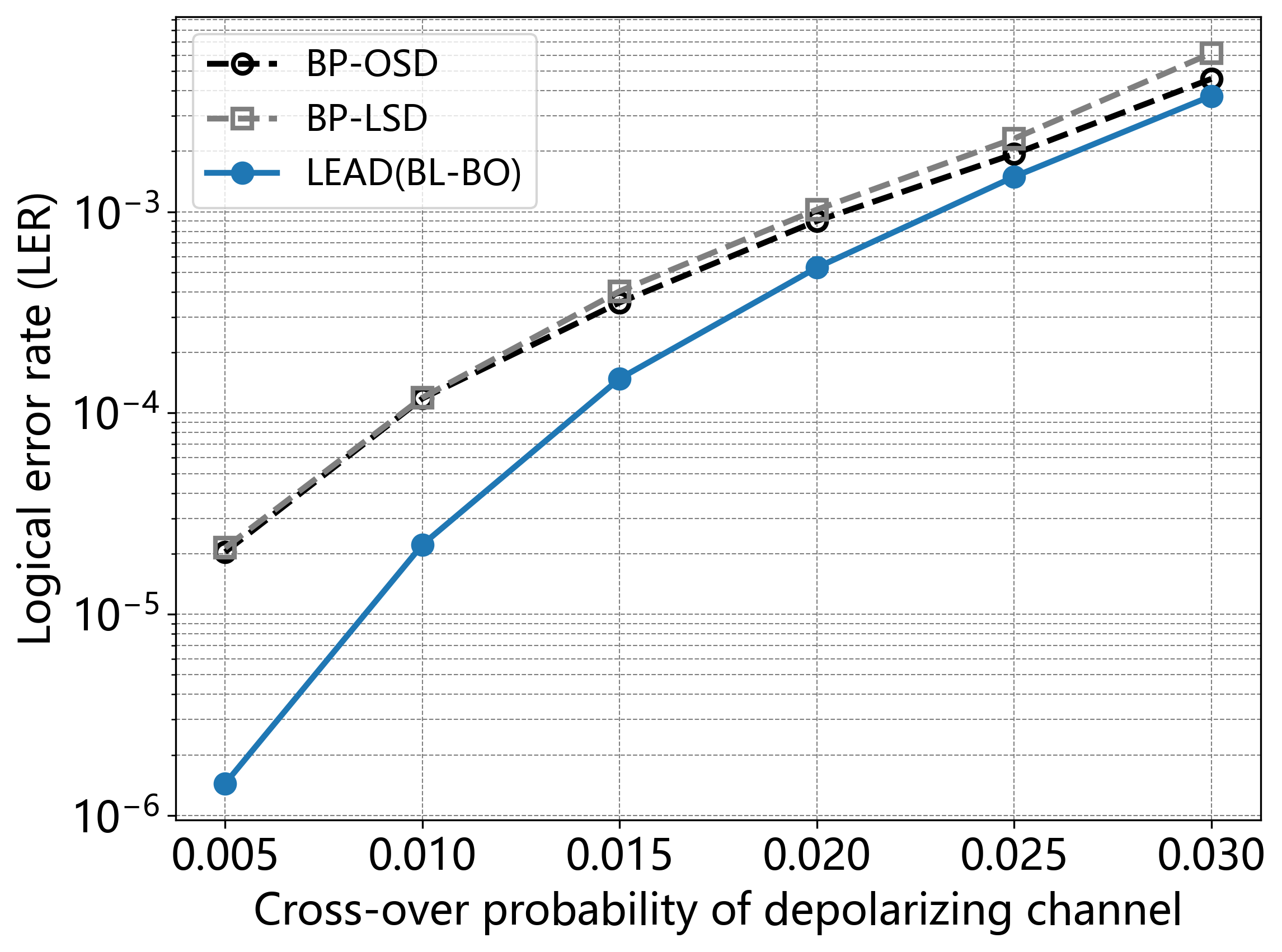}
        \label{fig:BCH(7,4)_test2}
    }
    \hfill
    \subfloat[]{
        \includegraphics[width=0.3\linewidth]{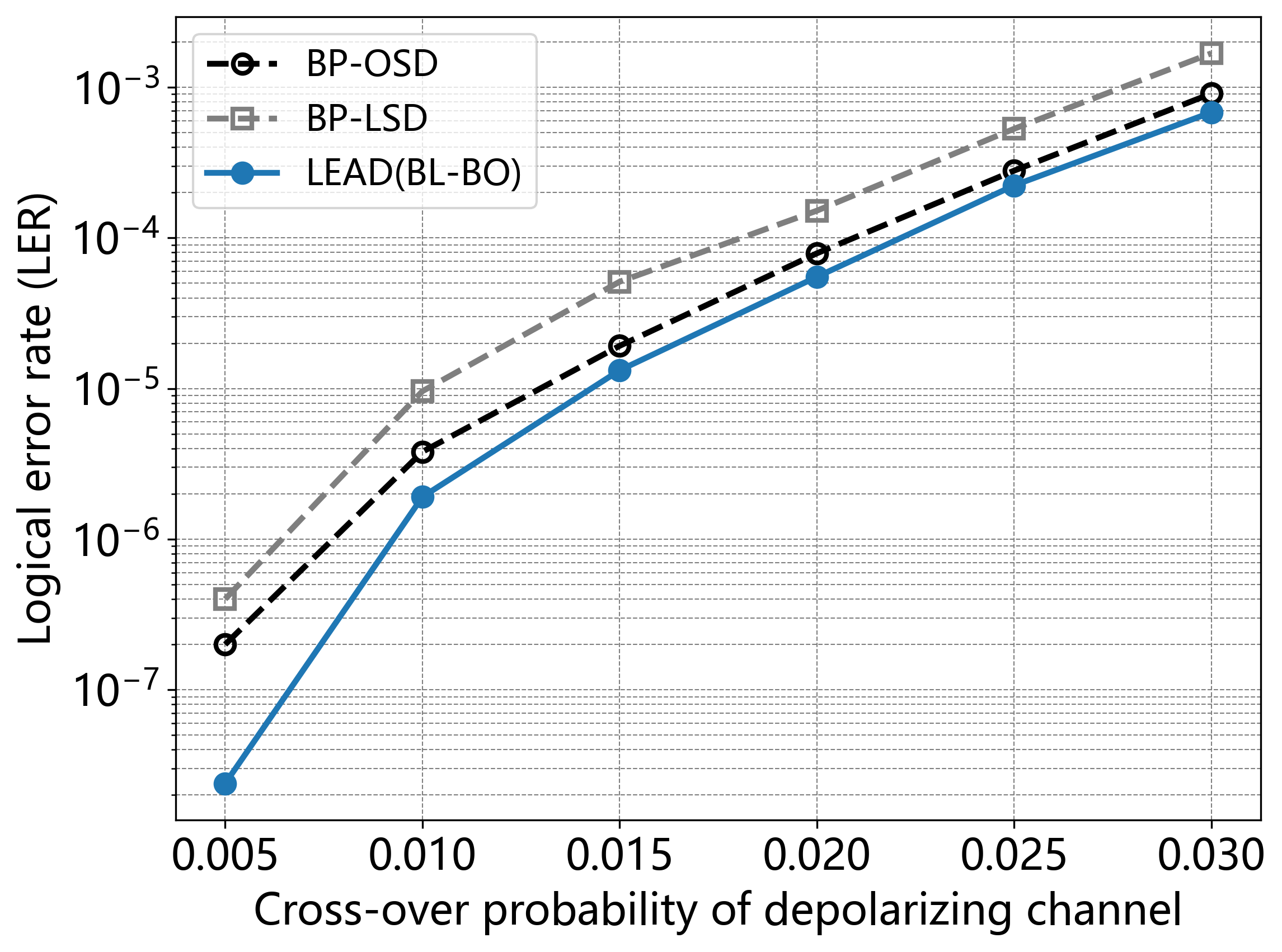}
        \label{fig:random}
    }

    \vspace{0.5cm}

    \subfloat[]{
        \includegraphics[width=0.3\linewidth]{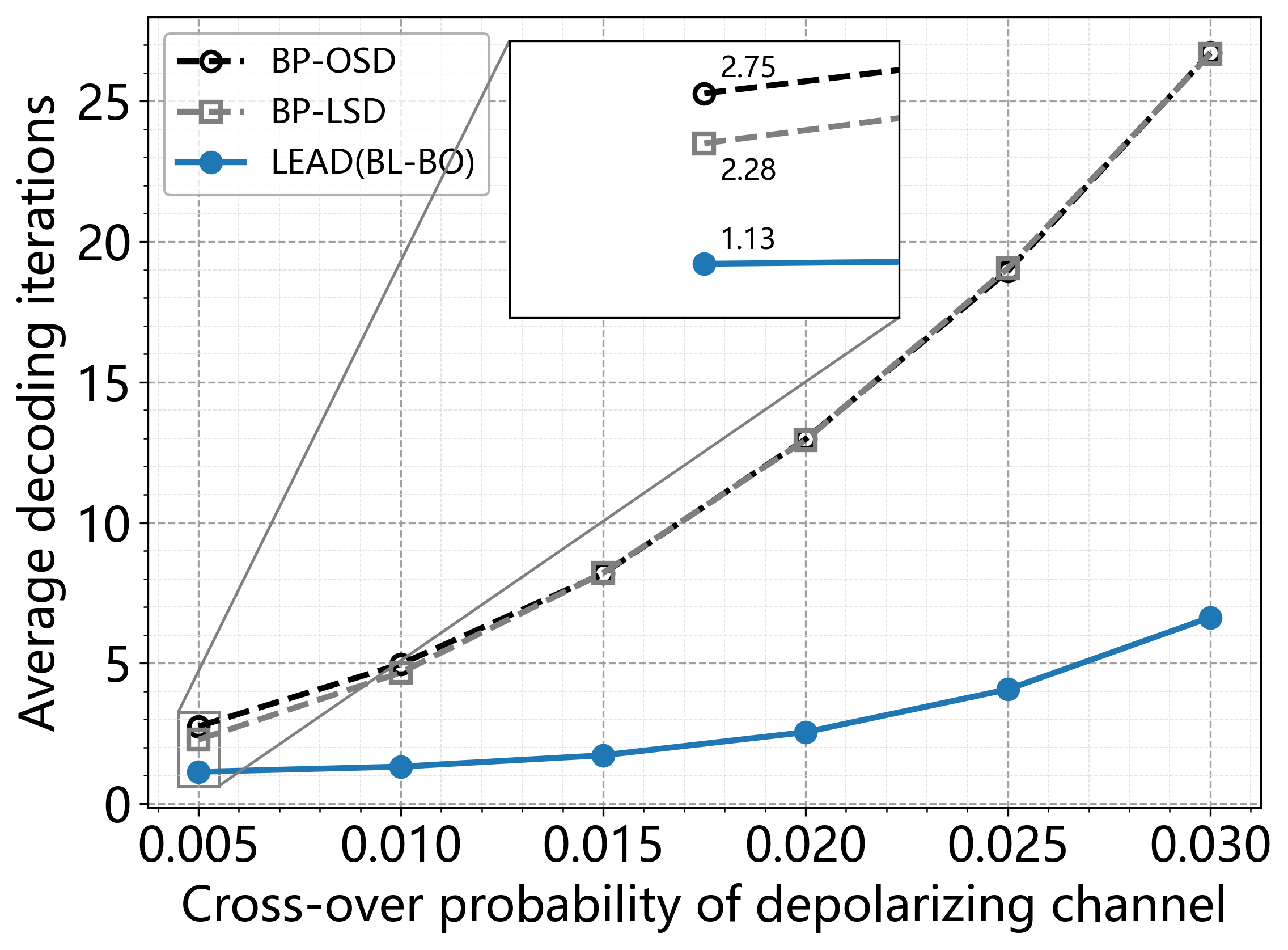}
        \label{fig:iteration2_Hamming(7,4)}
    }
    \hfill
    \subfloat[]{
        \includegraphics[width=0.3\linewidth]{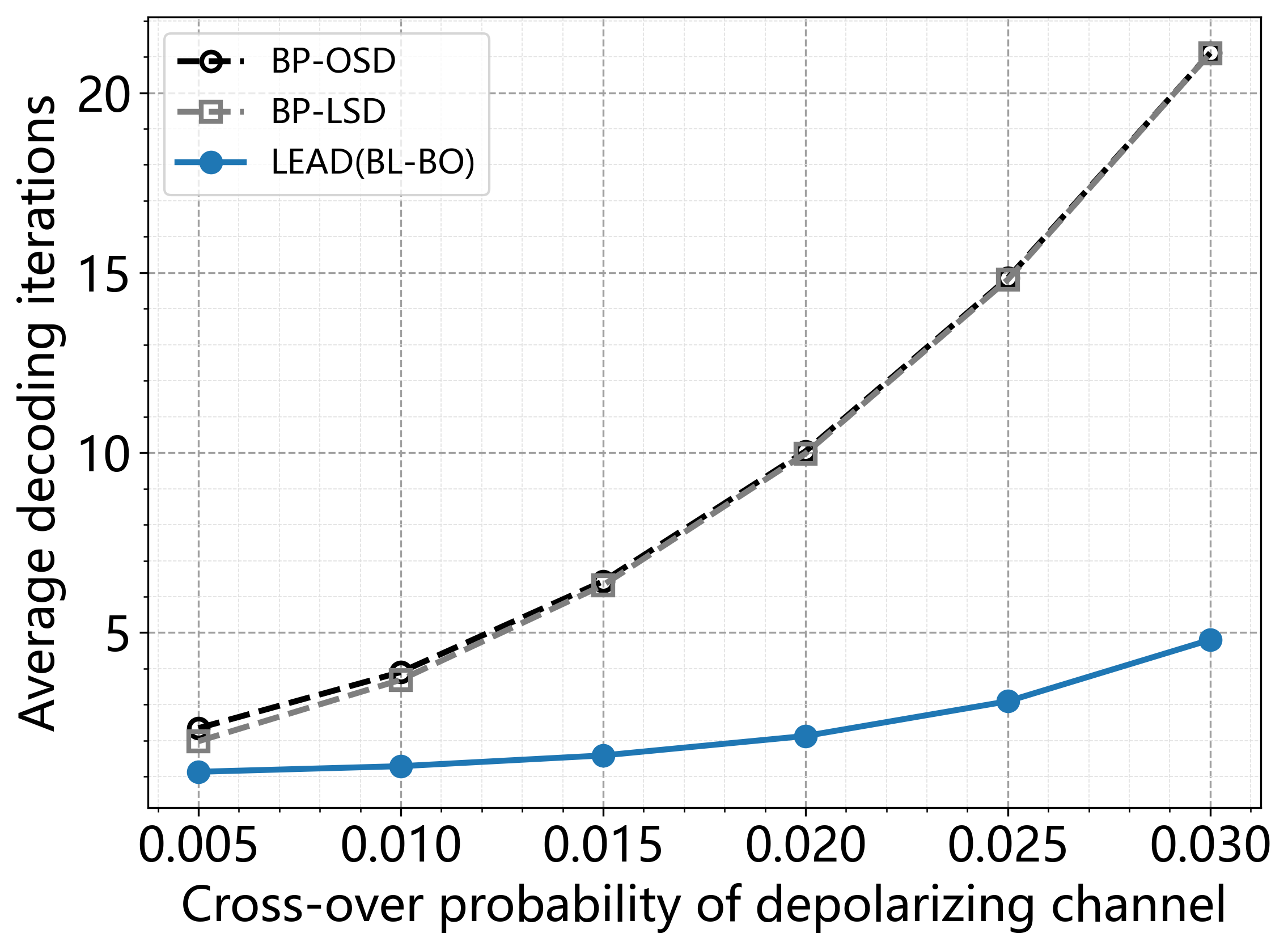}
        \label{fig:iteration_BCH(7,4)_test2}
    }
    \hfill
    \subfloat[]{
        \includegraphics[width=0.3\linewidth]{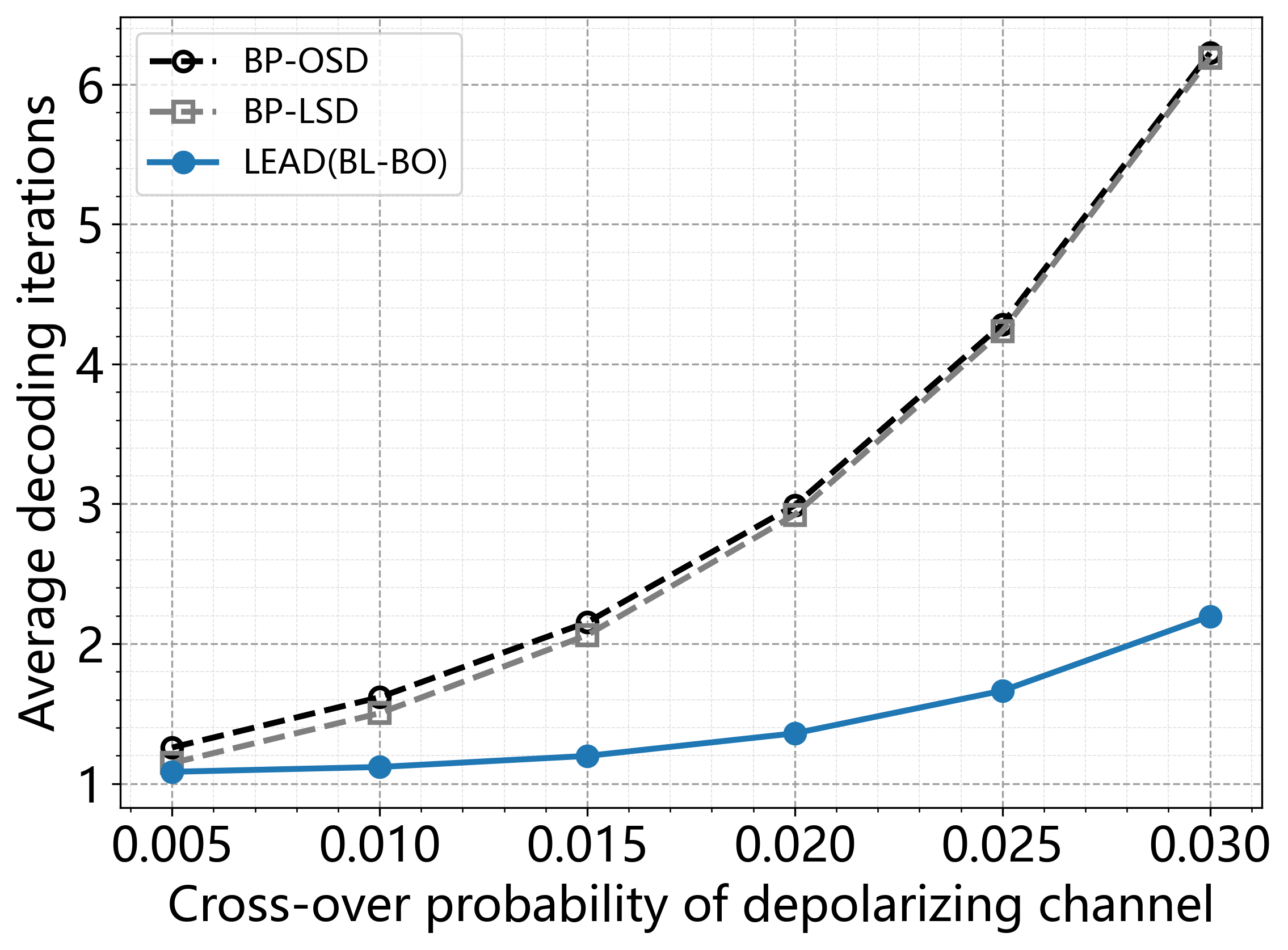}
        \label{fig:iteration_random}
    }

    \caption{Performance comparison of the LEAD framework against baseline decoders for various random quantum Tanner codes. Panels (a)--(c) illustrate the LER versus physical error probability $p$, and panels (d)--(f) present the corresponding average number of decoding iterations. The code parameters are: (a, d) $[\![392,48]\!]$, (b, e) $[\![392,8]\!]$, and (c, f) $[\![252,8]\!]$. Across all instances, the LEAD architecture (solid lines) consistently achieves superior error suppression and lower computational latency compared to the standalone BP--OSD and BP--LSD baselines.}
    \label{fig:regular}
\end{figure*}

We first consider a quantum Tanner code with parameters $[\![392,48]\!]$ \cite{perlin2023qldpc}.
Specifically, let $C_A$ be the $[7,4]$ Hamming code and $C_B$ its dual $C_B = C_A^\perp$.
Let $G$ be a cyclic group of order $8$,  from which two subsets
$A$ and $B$ are randomly selected such that $|A| = |B| = 7$.
This construction produces the target quantum Tanner code.

The LER results are shown in Fig.~\ref{fig:Hamming(7,4)}, and decoding iterations is presented in Fig.~\ref{fig:iteration2_Hamming(7,4)}.
Compared with the baseline BP--OSD and BP--LSD algorithms, the proposed LEAD algorithm achieves a lower LER and requires fewer decoding iterations.
For example, at $p=0.005$, the LEAD framework yields an LER of $1.7\times 10^{-5}$ , achieving a nearly two-order-of-magnitude improvement over the BP--OSD baseline ($7.2\times 10^{-4}$).
The efficiency gains are equally compelling: the LEAD algorithm achieves a dramatic $58.9\%$ reduction in average iterations (from $2.75$ to $1.13$) compared to standalone BP--OSD. Similarly, LEAD framework halves the computational burden of the BP--LSD baseline, lowering the iteration count from $2.28$ to $1.13$ (a $50.4\%$ decrease).


To evaluate the performance of the LEAD framework across different redundancy regimes, we extend our analysis to a second configuration: a $[\![392, 8]\!]$ quantum Tanner code, representing a shift in code rate from $R \approx 0.12$ to a more redundant $R \approx 0.02$.
Here, the local codes are a $[7,4]$ BCH code ($C_A$) and its dual $[7,3]$ code ($C_B = C_A^\perp$), with $G$ being a cyclic group of order $8$. The generating sets $A, B \subseteq G$ are randomly selected such that $|A| = |B| = 7$. As illustrated in Fig.~\ref{fig:BCH(7,4)_test2} and~\ref{fig:iteration_BCH(7,4)_test2}, the LEAD framework consistently achieves a concurrent reduction in both LER and average iteration count, reinforcing the performance gains observed in the previous case.

Next, we challenge the LEAD framework with non-standard local constraints by employing a randomly generated local code $C_A$. Defined by the parity-check matrix in Eq.~\eqref{eq:random_matrix}, this example serves to verify whether the performance gains are sustained when the local codes lack the structured algebraic properties of BCH or repetition codes.
\begin{equation}\label{eq:random_matrix}
 \mathbf{H}_{A} = \begin{bmatrix} 1 & 1 & 0 & 1 & 0 & 0 \\ 0 & 1 & 1 & 0 & 1 & 0 \\ 1 & 0 & 1 & 0 & 0 & 1 \end{bmatrix}.
\end{equation}
By taking $C_B = C_A^\perp$ and $G$ as a cyclic group of order $7$, we obtain a $[\![252, 8, d]\!]$ quantum Tanner code. The LER and iteration results are presented in Fig.~\ref{fig:random} and Fig.~\ref{fig:iteration_random}, respectively. At a target LER of $2.7 \times 10^{-7}$, the standalone BP--OSD baseline requires a physical error rate of $p \approx 0.005$, whereas the LEAD decoder maintains the same performance up to $p \approx 0.0078$—a $56\%$ improvement in error tolerance. This significant gain, coupled with a substantial reduction in average iterations (Fig.~\ref{fig:iteration_random}), further validates the versatility and error-correction prowess of the LEAD architecture even under unoptimized local structures.

\subsection{Generalizability Analysis and the Scaling Parameter}
\label{sec:Generalizability Analysis}
\begin{figure*}[t!]
    \centering
    \subfloat[]{
        \includegraphics[width=0.3\linewidth]{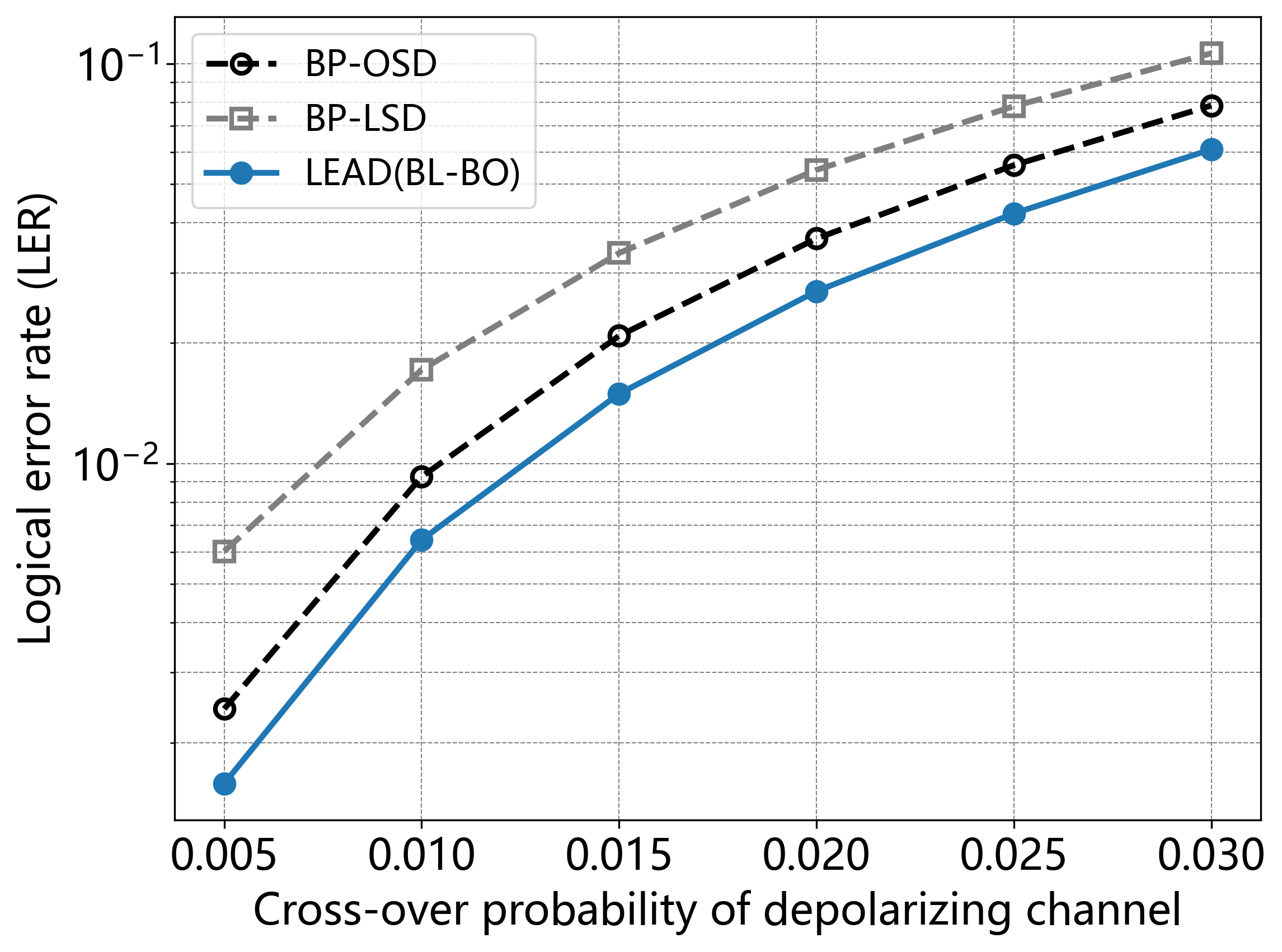}
        \label{fig:36_8_3}
    }
    \hfill
    \subfloat[]{
        \includegraphics[width=0.3\linewidth]{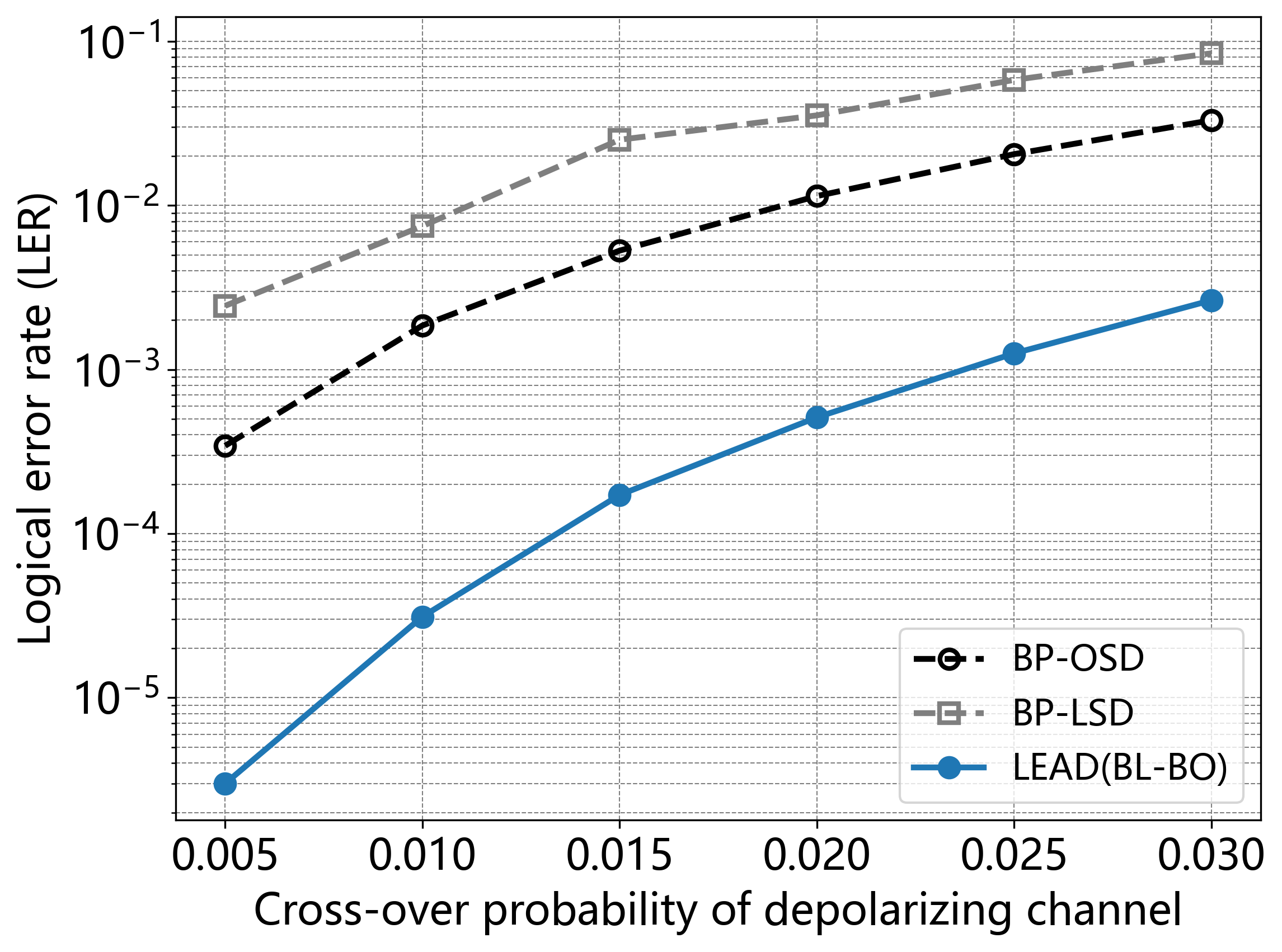}
        \label{fig:80_2_10}
    }
    \hfill
    \subfloat[]{
        \includegraphics[width=0.3\linewidth]{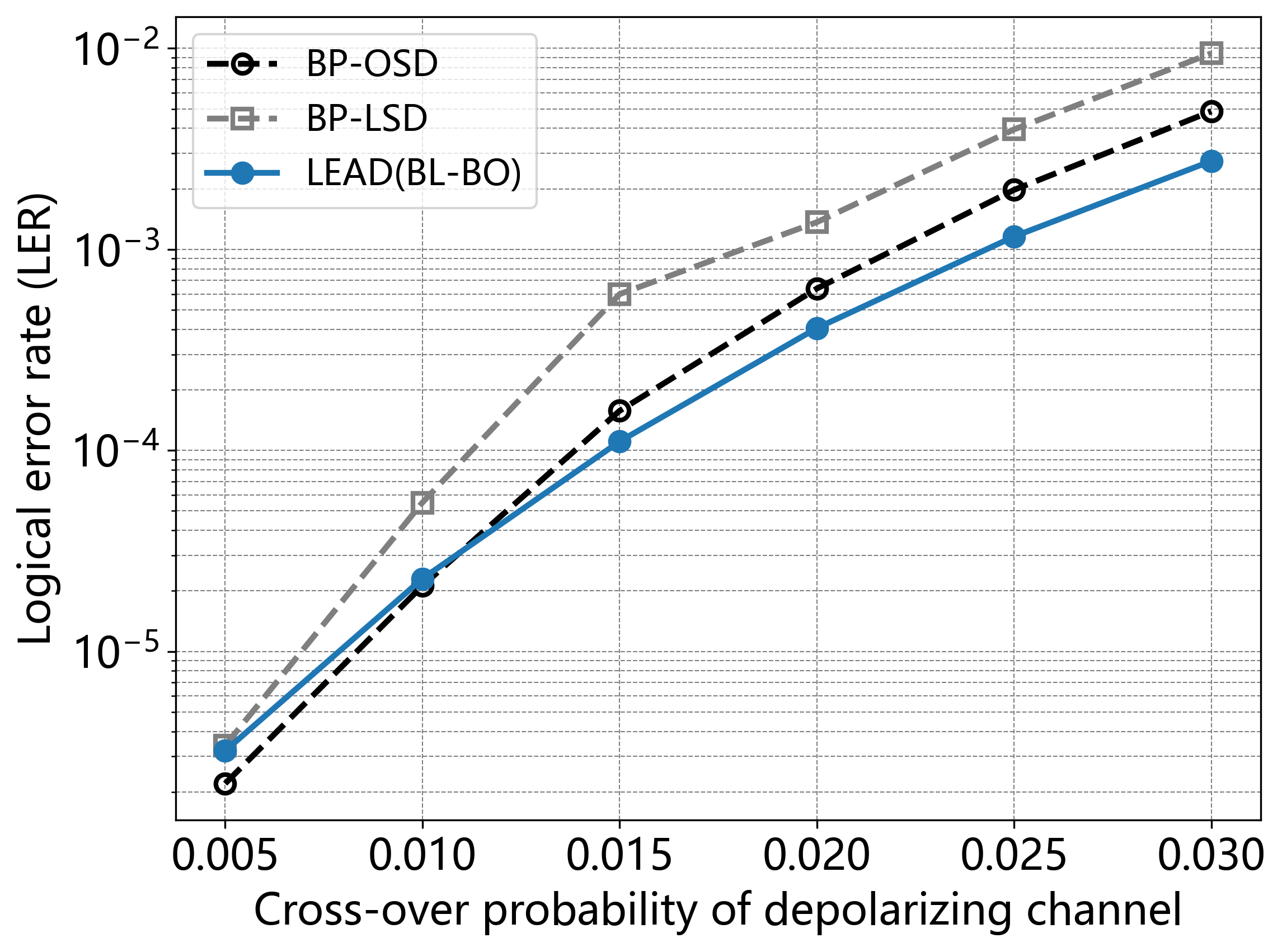}
        \label{fig:144_8_12}
    }

    \vspace{0.5cm}

    \subfloat[]{
        \includegraphics[width=0.3\linewidth]{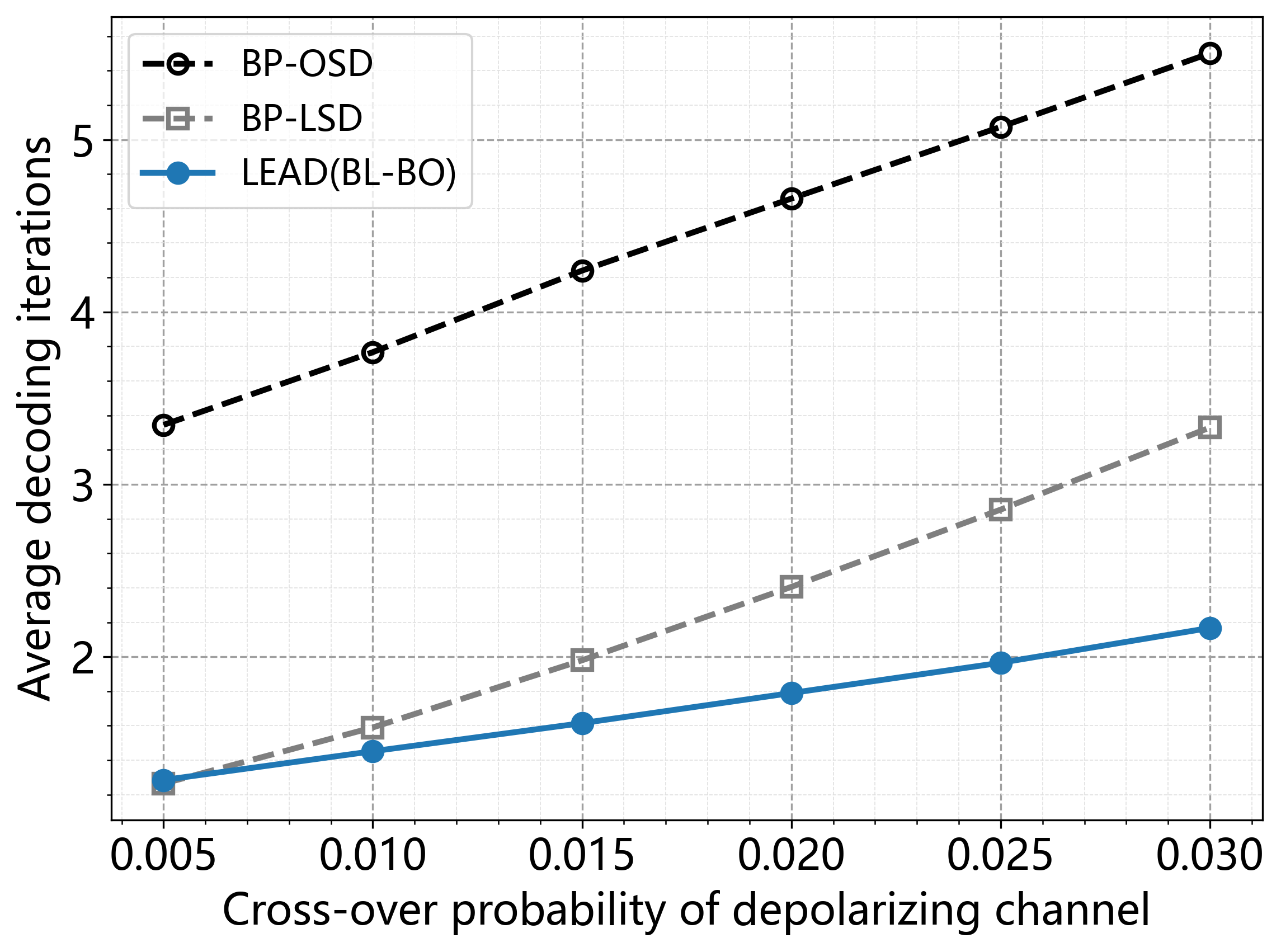}
        \label{fig:iteration_[[36_8_3]]}
    }
    \hfill
    \subfloat[]{
        \includegraphics[width=0.3\linewidth]{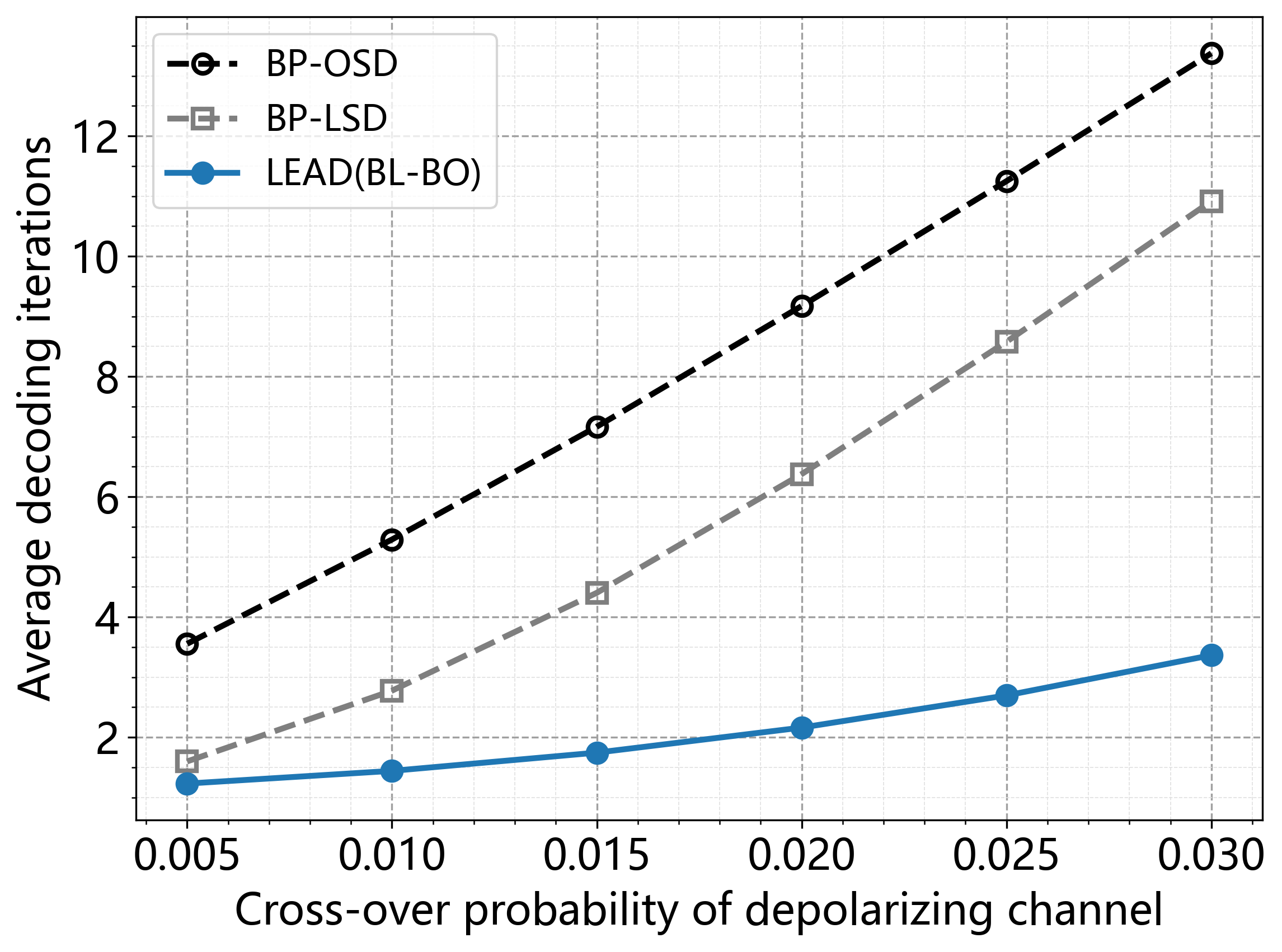}
        \label{fig:iteration_[[80_2_10]]}
    }
    \hfill
    \subfloat[]{
        \includegraphics[width=0.3\linewidth]{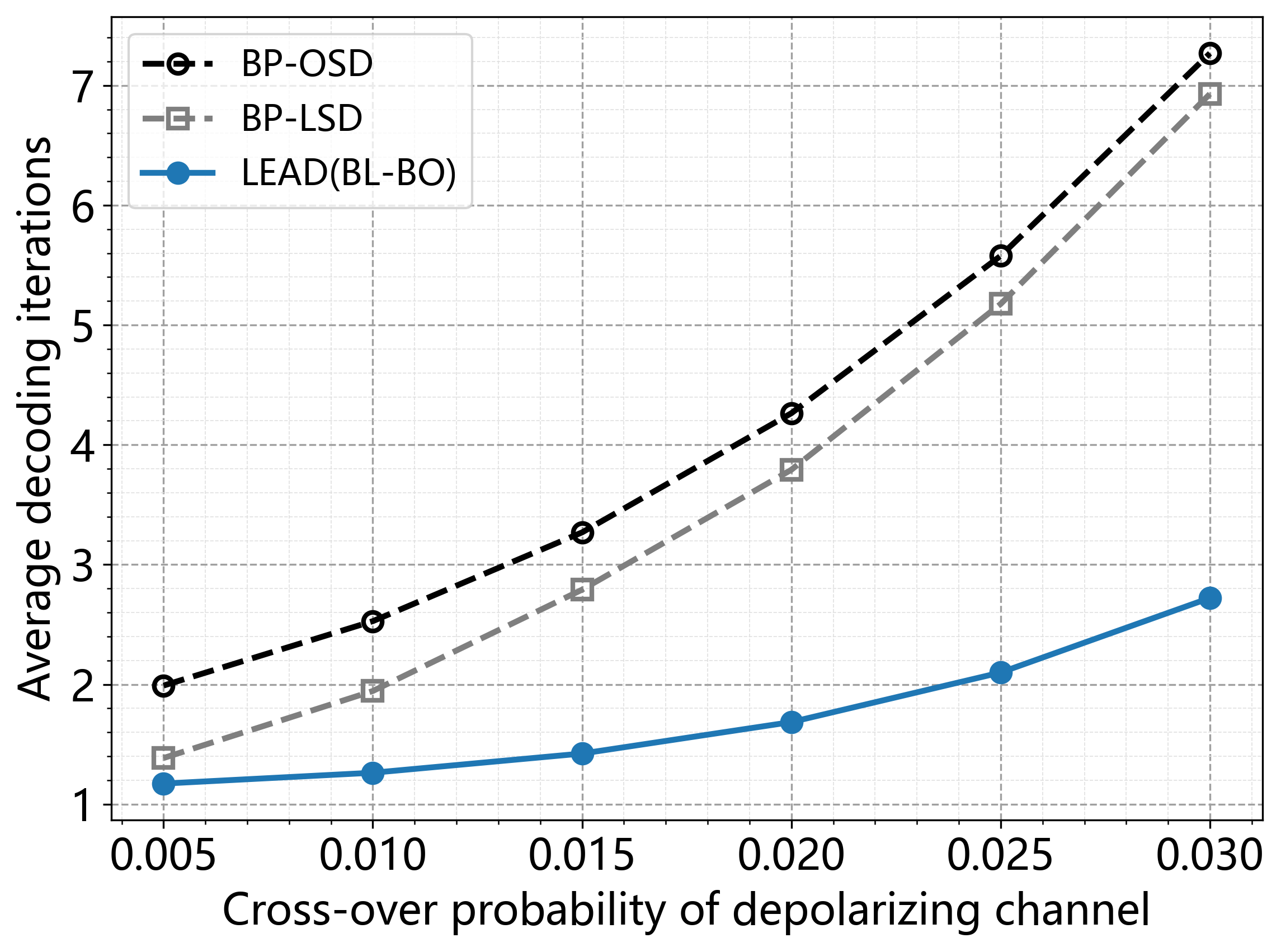}
        \label{fig:iteration_[[144_8_12]]}
    }

    \caption{Decoding performance comparison for various quantum Tanner codes derived from [2, 3]. Panels (a)–(c) present the LER as a function of physical error probability $p$, while (d)–(f) illustrate the corresponding average number of decoding iterations. The code parameters are: (a, d) $[\![36, 8, 3]\!]$, (b, e) $[\![80, 2, 10]\!]$, and (c, f) $[\![144, 8, 12]\!]$. Across all configurations, the LEAD framework (solid lines) consistently reduces decoding latency compared to standalone BP-LSD and  BP-OSD baselines, while maintaining or improving error-correction accuracy.}
    \label{fig:regular}
\end{figure*}

To further evaluate the generality of the proposed LEAD framework, we apply it to quantum Tanner codes constructed in previous works. In particular, we consider the $[\![36,8,3]\!]$ code constructed in \cite{radebold2025explicit}, as well as the $[\![80,2,10]\!]$ and $[\![144,8,12]\!]$ codes presented in \cite{leverrier2025small}. The consistent performance gains observed across codes from various construction families underscore the versatility of the LEAD framework, independent of the specific choice of the underlying finite group $G$ or the local constituent codes. The decoding performance is shown in Figs.~\ref{fig:36_8_3}--\ref{fig:iteration_[[144_8_12]]}.

While the LEAD framework generally improves decoding performance, we observe an anomalous behavior in the low-error regime for certain code instances. As illustrated in the performance curves (e.g., Fig. \ref{fig:144_8_12}), at a lower depolarizing error probability such as $p = 0.005$, the LEAD framework occasionally fails to provide a significant decoding gain in terms of LER over the standard BP-OSD baseline. While the LEAD framework consistently maintains a substantial advantage in decoding efficiency—requiring significantly fewer iterations (see Fig. \ref{fig:iteration_[[144_8_12]]})—this efficiency does not translate into improved error-correction accuracy in this specific regime. This discrepancy indicates that local overconfidence may be limiting the logical performance, presenting a critical bottleneck that warrants further investigation.

\begin{figure}[h]
    \centering
    \includegraphics[width=0.9\linewidth]{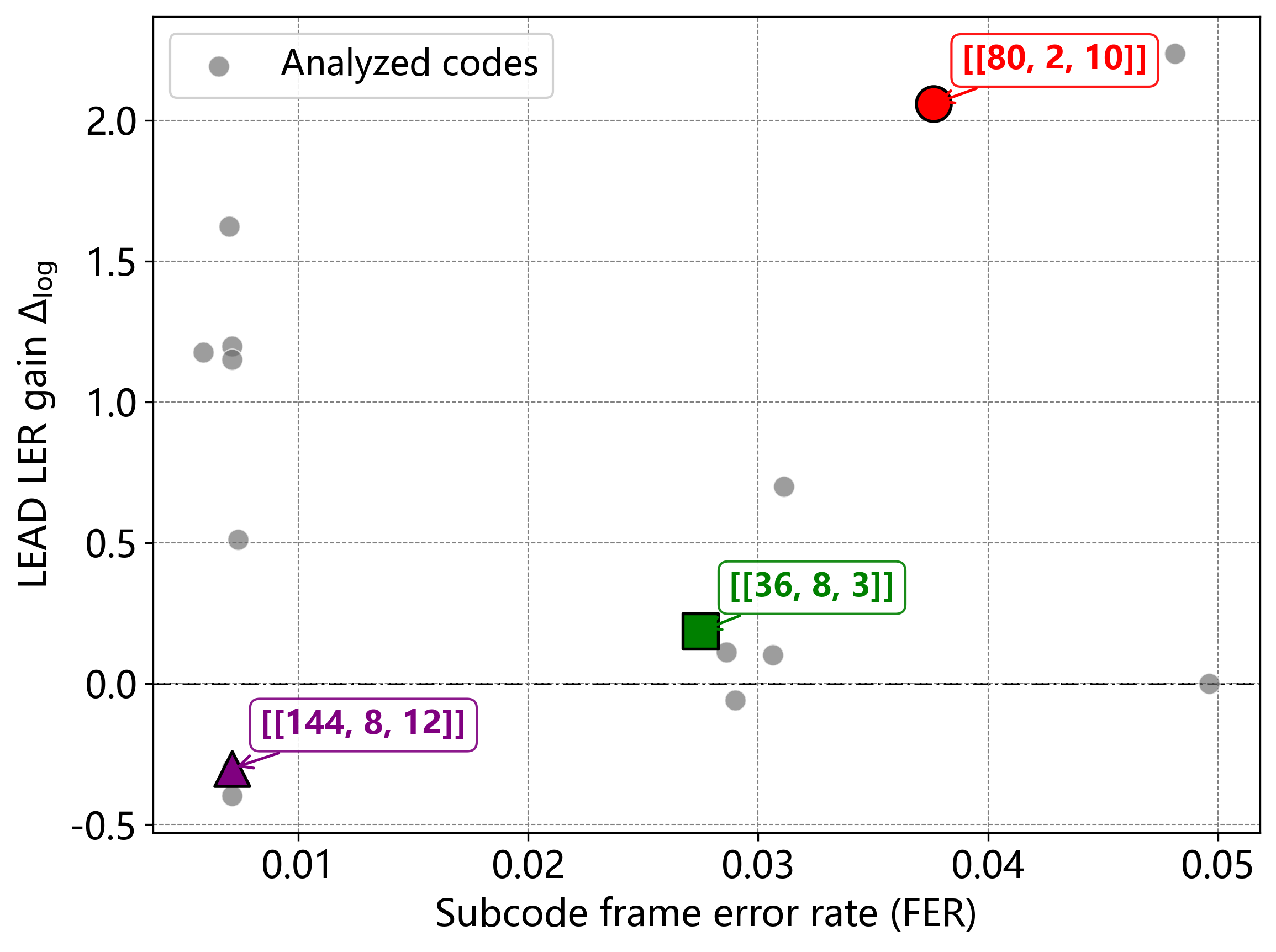}
    \caption{Analysis of LEAD decoding gain ($\Delta_{\log}$) vs. subcode FER for different quantum Tanner code instances at $p=0.005$. The horizontal dash-dotted line at $0$ denotes the baseline; points below it indicate performance degradation. The results illustrate that local decoding accuracy alone is insufficient to predict global performance improvement, highlighting that local overconfidence—rather than raw subcode error rate—is the primary driver of performance bottleneck in the global belief propagation process.
}
    \label{fig:analyzed_code}
\end{figure}

To investigate the root cause of this performance bottleneck, we systematically evaluate the decoding gain across a diverse set of quantum Tanner codes at $p = 0.005$. We quantify the LER gain using the logarithmic difference:

\[
\Delta_{\log} = \log_{10}(\mathrm{LER}_{\mathrm{BP\!-\!OSD}})
- \log_{10}(\mathrm{LER}_{\mathrm{LEAD(BL-\!BO)}})
\]

where a positive $\Delta_{\log}$ indicates improvement. As shown in the scatter plot in Fig.~\ref{fig:analyzed_code}, the results reveal a counterintuitive decoupled relationship between local performance and global gain: a lower subcode frame error rate (FER) does not strictly guarantee a higher global decoding gain. For instance, the $[\![144, 8, 12]\!]$ instance exhibits a low subcode FER but suffers from a slight negative gain ($\Delta_{\log} < 0$), suggesting that raw local accuracy is not the sole metric for assessing the contribution of a local decoder.

This bottleneck can be attributed to the interplay between local code distance and the expansion properties of the underlying Tanner complex. When the local distance is limited, a local decoder in the low-noise regime is susceptible to misconvergence, erroneously identifying low-weight error patterns as valid local codewords and outputting highly polarized but incorrect probability distributions. If the expansion of the complex is insufficient to distribute these errors, these overconfident misjudgments act as strong noise sources. Within the belief propagation framework, this triggers a positive feedback loop of erroneous soft information, trapping the global decoder in sub-optimal local minima.

\begin{figure}[h]
    \centering
    \includegraphics[width=0.9\linewidth]{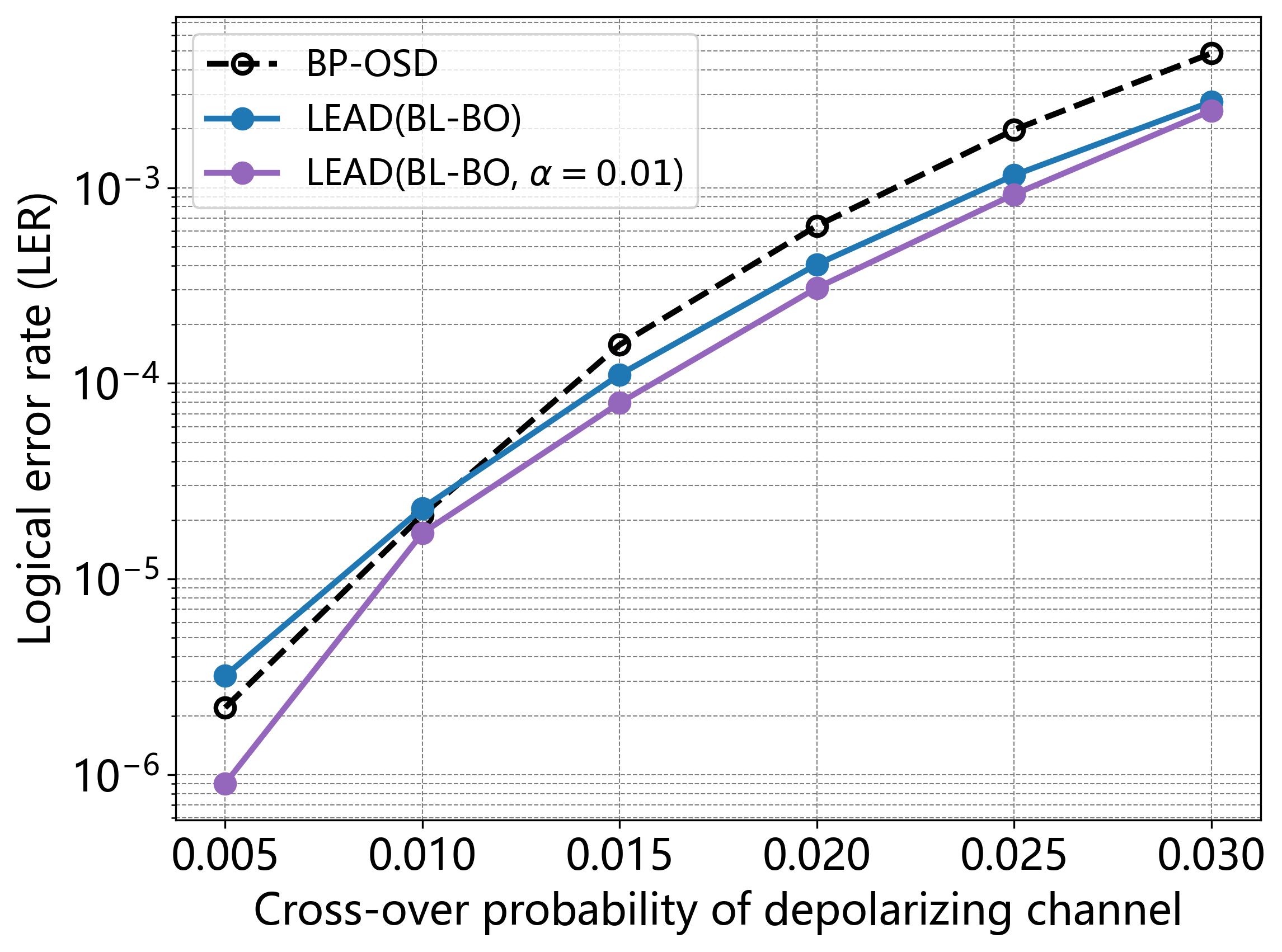}
   \caption{Impact of the scaling parameter $\alpha$ on the decoding performance of the $[\![144,8,12]\!]$ quantum Tanner code. While the unscaled configuration ($\alpha=1.0$) suffers from performance degradation in the low-error regime due to local overconfidence, applying strong dampening ($\alpha=0.01$) effectively regularizes the local soft information. This transition demonstrates how the proposed soft-thresholding mechanism prevents erroneous local beliefs from dominating the global belief propagation process, thereby restoring the LER advantage over the BP-OSD baseline.}
    \label{fig:compare_alpha}
\end{figure}

The introduction of the scaling parameter $\alpha$ effectively acts as an information-theoretic regularization. To address the observed performance bottleneck in specific code instances, we leverage this mechanism to dampen extreme local beliefs, thereby allowing global parity constraints to resolve local inconsistencies. As demonstrated in Fig.~\ref{fig:compare_alpha} for the $[\![144, 8, 12]\!]$ code, while the unscaled configuration ($\alpha=1.0$) exhibits noticeable degradation at low physical error rates, applying strong dampening ($\alpha=0.01$) effectively eliminates this bottleneck and restores superior performance across the entire noise regime.

Mathematically, $\alpha$ functions as a soft-thresholding mechanism that regularizes the highly polarized distributions produced by local decoders. By aggressively dampening local soft information, LEAD filters out high-confidence outliers, ensuring that global constraints can resolve local inconsistencies. This shifts the focus from maximizing raw local accuracy to ensuring the reliability and consistency of information passed across hierarchical layers, enabling robust global convergence even when local decoders face distance limitations.

This behavior naturally explains the decoder preference observed in our experiments. Although OSD is theoretically stronger in terms of local search capability, employing OSD at the local level often produces excessively polarized reliability estimates that can dominate subsequent belief propagation updates once converging to an incorrect local codeword. In contrast, LSD provides softer and better-calibrated information, allowing global parity constraints to progressively correct local inconsistencies. Consequently, employing BP-LSD as the local decoder and BP-OSD as the global decoder achieves a better balance between local reliability and global convergence, while also reducing the average decoding iterations observed in our experiments.

\section{Discussion and Future Work}
\label{sec:discussion}
\subsection{Relationship with Recent Concurrent Works}
During the finalization of this manuscript, we became aware of two recent independent works on the decoding of quantum Tanner codes: the generalized check node approach by Mostad et al.~\cite{Lin2026}, and the SOGRAND framework by Rapp et al.~\cite{Ken2026} While these studies were developed concurrently with our work, they address different facets of the decoding challenge.

While LEAD focuses on the high-level scheduling architecture and the convergence logic of local error amendments, the aforementioned works offer significant advancements in local constraint optimization and probabilistic soft-information processing, respectively. We highlight that LEAD is fundamentally compatible and complementary with these new methodologies:

\begin{itemize}
  \item Architectural Integration: The linear-time scheduling of LEAD can naturally incorporate Generalized Check Nodes as high-performance local operators to enhance the error suppression threshold without sacrificing scalability.
  \item Synergy for Real-time Decoding: SOGRAND demonstrates the power of soft-output guessing in mitigating trapping sets. LEAD provides the necessary structural framework to deploy such sophisticated algorithms in practical quantum hardware.
\end{itemize}

\subsection{ML-Based Parameter Optimization}
The effectiveness of the LEAD framework, as shown in our simulations, relies on the proper regularization of local beliefs through the scaling parameter $\alpha$. Our findings indicate that $\alpha$ acts as a critical \emph{soft-thresholding } mechanism to prevent the local overconfidence that otherwise traps global belief propagation. Currently, $\alpha$ is treated as a static hyperparameter; however, its optimal value may depend on the specific code instance and the instantaneous noise characteristics.

A promising direction for future research is the integration of Graph Neural Networks (GNNs) to achieve autonomous, real-time parameter tuning. Given that the decoding process is inherently defined on a graph-based Tanner complex, a lightweight GNN could be trained to perceive the syndrome distribution and predict node-specific or iteration-dependent values of $\alpha$. Such a data-driven approach would eliminate the need for heuristic tuning and potentially allow the decoder to adaptively balance local assistance with global consistency. Together with the parallel developments in local decoding operators, these advancements pave the way toward a unified and intelligent ecosystem for efficient, real-time quantum error correction.

\section{Conclusion}\label{sec:conclusion}
In this paper, we have proposed a decoding framework for quantum Tanner codes, referred to as the LEAD algorithm. The algorithm first leverages the structural decomposability of the Cayley complex to partition the global quantum code into local subcodes defined on the $Q$-neighborhoods of complex vertices, performing parallel estimation of error probabilities on these local views. Subsequently, by exploiting the topological symmetry of the Cayley complex and introducing a soft-information regularization mechanism to mitigate local overconfidence, these local estimates are effectively aggregated into a global error probability prior. Compared with generic BP-based decoding algorithms, LEAD explicitly exploits the local code structure inherent in quantum Tanner codes; and compared with existing local-decoder-based approaches, it avoids computationally expensive mismatch-vector decomposition, thereby significantly improving practicality.

We evaluated the LEAD framework across several quantum Tanner code instances with different code lengths and rates via Monte Carlo simulations. The results demonstrate that LEAD significantly outperforms the standard BP--OSD and BP-LSD baselines in terms of LER, while simultaneously achieving a substantial reduction in the average number of decoding iterations. Crucially, our investigation into heterogeneous configurations reveals that the LEAD framework achieves an optimal balance between accuracy and efficiency when employing a lightweight LSD as the local sub-decoder and a powerful OSD for global post-processing. This finding provides a specific and valuable reference for future hardware implementations of high-throughput quantum decoders.

By striking a highly favorable balance between decoding performance and computational complexity, the LEAD framework aligns well with the stringent requirements of early fault-tolerant quantum computing architectures, where physical qubit resources are limited and gate fidelities remain imperfect. Ultimately, it provides a highly practical and scalable decoding solution for robust operations on future quantum processors.

\appendices

\section{Illustrative Example of the LEAD Decoder}
\label{append:A}
We illustrate the decoding procedure with a simplified example.
Let $C_A$ and $C_B$ be the three-bit repetition code and its dual code, respectively, and let $G$ be a cyclic group of order $4$.
By randomly selecting generating subsets $A$ and $B$ from $G$ such that $|A| = |B| = 3$, one can construct a quantum Tanner code with parameters $[\![36,8]\!]$ \cite{perlin2023qldpc}. The corresponding $X$-type parity-check matrix is shown in Eq.~\eqref{eq:matrix}, where the highlighted blocks correspond to the local parity-check matrices associated with three different vertices $v_1, v_2, v_3 \in V$.
The local parity-check matrix is given in Eq.~\eqref{eq:product}, derived from a $[9,7,2]$ classical product code, which can only correct a limited set of errors.

\begin{equation}
\label{eq:matrix}
\resizebox{\linewidth}{!}{$
\left[
\begin{array}{ccccccccccc}
 \mathbf{H}_{1} &
\multicolumn{9}{c}{
\boxed{
\begin{array}{*{9}{c}}
X_1 & X_2 & I_3 & X_4 & X_5 & I_6 & X_7 & X_8 & I_9 \\
I_1 & X_2 & X_3 & I_4 & X_5 & X_6 & I_7 & X_8 & X_9
\end{array}
}
} & \cdots
\\[6pt]
\cdots &
\multicolumn{9}{c}{
\boxed{
\begin{array}{*{9}{c}}
X_{1} & I_{6} & X_{8} & X_{10} & I_{11} & X_{12} & X_{13} & X_{14} & I_{15} \\
X_{1} & X_{6} & I_{8} & X_{10} & X_{11} & I_{12} & X_{13} & I_{14} & X_{15}
\end{array}
}
} &  \mathbf{H}_{2}
\\[6pt]
\cdots &
\multicolumn{9}{c}{
\boxed{
\begin{array}{*{9}{c}}
I_{11} & X_{13} & X_{16} & X_{17} & I_{18} & X_{19} & I_{20} & X_{21} & X_{22} \\
X_{11} & I_{13} & I_{16} & X_{17} & X_{18} & X_{19} & X_{20} & I_{21} & X_{22}
\end{array}
}
} & \mathbf{H}_{3}
\\[6pt]
\vdots & \vdots & \vdots & \vdots & \vdots & \vdots & \vdots & \vdots & \vdots & \vdots & \vdots
\end{array}
\right]
$}
\end{equation}

\begin{equation}
\label{eq:product}
 \begin{bmatrix}
X & I & X & X & I & X & X & I & X \\
I & X & X & I & X & X & I & X & X
\end{bmatrix}
\end{equation}

In this instance, the local codes possess weak error-correction capability, making them highly susceptible to mis-decoding. Traditional algorithms that strictly require all local decoders to succeed, such as the approach in \cite{leverrier2023decoding}, are prone to failure in this regime. In contrast, the proposed LEAD algorithm only requires a subset of local codes to yield meaningful soft information to effectively exploit the local structural topology.
\begin{enumerate}

\item \textbf{Initialization:} The global prior probability vector $\hat{\mathbf{p}} \in \mathbb{R}^{36}$ is initialized to $\mathbf{0}_{36}$.

\item \textbf{Local Decoding \& Confidence Boosting (Phase 1):} Assume a single phase-flip error $Z_1$ occurs. This qubit is shared by vertices $v_1$ and $v_2$. The local decoders produce estimates $\hat{\mathbf{p}}_{v1}$ and $\hat{\mathbf{p}}_{v2}$, where
     \begin{multline}
\hat{\mathbf{p}}_{v1} = \big[
    0.21_1,\; 0.12_2,\; 0.04_3,\; 0.50_4, \\
    0.12_5,\; 0.04_6,\; 0.21_7,\; 0.12_8,\; 0.04_9
\big],
\end{multline}
and
\begin{multline}
\hat{\mathbf{p}}_{v2} = \big[
    0.25_1,\; 0.10_6,\; 0.10_8,\; 0.25_{10}, \\
    0.10_{11},\; 0.10_{12},\; 0.50_{13},\; 0.10_{14},\; 0.10_{15}
\big].
\end{multline}
     Note that due to the weak local distance, the local decoders may exhibit overconfidence in incorrect positions (e.g., the $0.50$ peaks at indices 4 and 13), while providing only moderate evidence for the true error at index 1.

\item \textbf{Soft Information Regularization (Phase 2):} We aggregate the soft information across all views. For qubit $i=1$, the set of local estimates is $\mathcal{S}_1 = \{0.21, 0.25\}$. The regularized global probability is computed as:
\begin{equation}
\hat{p}_1 = \alpha \cdot \text{Average}(\mathcal{S}_1) = 1 \times \frac{0.21 + 0.25}{2} = 0.23.
\end{equation}
This averaging and scaling process (controlled by $\alpha$) dampens erroneous local peaks and concentrates probability mass on qubits supported by multiple local views.

\item \textbf{Global Decoding (Phase 3):} The aggregated vector $\hat{\mathbf{p}}$ is fed into the global decoder. Even with a moderate initial probability (0.23), the global BP-OSD utilizes the full parity-check constraints to resolve the remaining ambiguity and converge to the correct error $\hat{\mathbf{e}}$.
\end{enumerate}

\bibliographystyle{unsrt}
\bibliography{sample}



\end{document}